\documentclass[journal]{IEEEtran}

\usepackage{amsmath,amsfonts}
\usepackage{amsthm}
\usepackage{color}
\usepackage{booktabs}
\definecolor{light-blue}{rgb}{0.6,0.6,1}
\usepackage[table]{xcolor}
\usepackage{subfigure}
\usepackage{url}
\usepackage{verbatim}
\usepackage{graphicx}
\usepackage{epstopdf} 
\usepackage{cite}
\usepackage{xcolor}
\usepackage{amssymb}

\newcommand{\expect}[1]{\mathbb{E}\left[#1\right]}

\newcommand{\var}{\sigma^2}  

\newcommand{\Nsc}{N_{\rm sc}} 
\newcommand{\Nfft}{N_{\rm fft}} 
\newcommand{\Ncp}{N_{\rm cp}}  

\newcommand{\Ninfo}{N_{\rm info}} 
\newcommand{\Nos}{N_{\rm OS}} 
\newcommand{\snr}{\mathsf{snr}} 
\newcommand{\SINR}{\mathsf{SINR}} 
\newcommand{\ICI}{\mathsf{ICI}} 
\newcommand{\jrm}{{\rm j}}
\newcommand{\Ical}{\mathcal{I}} 
\newcommand{\Xcal}{\mathcal{X}}

\newtheorem{Cor}{Corollary}

\newtheorem{Prop}{Proposition}
\newtheorem{Lem}{Lemma}
\newtheorem{Rem}{Remark}

\begin{document}

\title{Repeated-and-Offset QPSK for Low-PAPR DFT-s-OFDM in Satellite Communications}
\author{\IEEEauthorblockN{Renaud-Alexandre Pitaval}
\thanks{Renaud-Alexandre Pitaval is with Shannon Research Center, Huawei Technologies Sweden AB, Kista 164 94, Sweden (e-mail: renaud.alexandre.pitaval@huawei.com).
}}

%
%
%

\maketitle

\begin{abstract}
Motivated by the convergence of terrestrial cellular networks and satellite communications, this article considers an adaptation of offset quadrature phase shift keying (OQPSK),  traditionally used with single-carrier waveforms in satellite systems, to discrete Fourier transform spread  orthogonal frequency-division multiplexed (DFT-s-OFDM), as employed in the uplink of terrestrial systems. 
First, analytical signal-to-interference-plus-noise (SINR) expressions are derived for DFT-s-OFDM  with frequency-domain spectral shaping (FDSS) carrying independently distributed $\pi/2$-BPSK or QAM symbols and received with single-tap equalization,  as  in 5G. 
Next, a correlation-induced spectral shaping technique, termed \emph{repeated-and-offset QPSK (RO-QPSK)}, is introduced,  relying solely on bit-level processing prior to conventional QPSK modulation. Specifically, the input bits are  Manchester encoded (repeated and flipped) with an offset between the in-phase and quadrature branches, resulting in order-one OQPSK-like  modulation. 
The  induced correlation between  consecutive QPSK symbols produces a Hann-shaped transmit spectrum   
yielding a peak-to-average power ratio (PAPR) on the order of 2 dB without explicit FDSS.
At the receiver, the repetition structure is exploited through post-DFT-despreading symbol combining, and the corresponding end-to-end SINR with this transmitter-receiver pair is derived in closed form.      
Theoretical analysis and simulation results show that RO-QPSK provides performance gains in narrowband and moderately frequency-selective channels, as encountered in satellite communications, while remaining on par with conventional 5G  uplink schemes in other scenarios.   
The combination of RO-QPSK with  FDSS  is also investigated, enabling further PAPR reduction while maintaining comparable link-level performance.   
\end{abstract}


\section{Introduction}
Satellite communications are expected to be natively supported in 6G,  
providing wide-area coverage and ubiquitous connectivity,  
and enabling emerging applications such as remote internet-of-things (IoT). 
Support for non-terrestrial network (NTN)  connectivity has already been investigated and specified within 5G.  
In Rel. 15 and Rel. 16, 3GPP demonstrated~\cite{3GPP_TR_38811,3GPP_TR_38821} that handheld user equipments (UEs) (i.e. commercial smartphones) can connect to both terrestrial networks (TNs) and NTNs, 
thereby extending satellite access beyond dedicated satellite terminals with higher power class and larger antenna apertures, such as very-small-aperture terminals (VSATs) used in traditional satellite systems~\cite{Panaitopol22}. 
  
A key outcome of these studies is that satellite connectivity
can be supported with the existing 5G NR waveform, namely  orthogonal frequency division multiplexing (OFDM)  for downlink (DL) and discrete Fourier transform spread  OFDM (DFT-s-OFDM) for uplink (UL), without requiring a dedicated satellite waveform. 
As a result, the 5G NR air interface can be largely reused to provide connectivity across both TN and NTN.   
However, direct satellite access from handheld devices primarily targets low-data-rate services and voice over IP (VoIP), and remains challenging under unfavorable propagation conditions. 
Even for such services,   NTN coverage enhancement techniques were introduced in 3GPP Rel. 18  to support handheld UEs at satellite beam edges.  

\subsection{Satellite Channels} 
In satellite communications, the limited link budget~\cite{ErkaiPIMRC24} requires often both UEs and satellites to operate close to the saturation region of their power amplifiers~\cite{SaarnisaariL21}.  
Therefore, as has long been the case for traditional satellite operators, signals with low peak-to-average power ratio (PAPR) are  
of particular importance to ensure power efficiency and avoid nonlinear distortion. 

Compared to terrestrial channels, satellite channels typically consist of only a few resolvable paths, often dominated by line-of-sight (LOS) conditions.  
Consequently, NTN channels generally exhibit very little frequency selectivity, and  many satellite channel models rely on frequency-flat fading assumptions~\cite{3GPP_TR_38811}.  
In addition, satellite links are characterized by large Doppler shifts and long propagation delays.  

In practical systems, however, Doppler effects can be largely compensated~\cite{QC_WiSEE23,orellana2026capacityboundsdopplerofdm}.
The NTN extension in 5G primarily targets UEs with Global Navigation Satellite System (GNSS) capabilities, enabling pre-compensation  in both UL and DL~\cite{3GPP_TR_38811,QC_WiSEE23}. This significantly reduces frequency offsets  and timing misalignments, allowing the reuse of 5G NR synchronization signals in NTN scenarios.  
As a result, the residual frequency offset during data transmission is typically moderate and can be further tracked and compensated, for example, through phase-tracking mechanisms. 
In the absence of GNSS capability at the UE, which is currently under investigation for 5G-Advanced (Rel. 20),  uplink frequency synchronization becomes more challenging. Nevertheless, it can be achieved  through enhanced network assistance~\cite{R1-2506612} and/or Doppler-resilient random access preambles~\cite{BerggrenComLet25}. Once initial access is established,  Doppler effect can again be largely compensated during data transmission.

\subsection{3GPP Low-PAPR Waveform and Modulation}
3GPP standards since 4G LTE Rel. 8 are based on OFDM combined with quadrature amplitude modulation (QAM) for data-transmission. A well-known drawback of OFDM is its high PAPR, which requires significant power back-off at the transmitter to avoid signal distortion due to power amplifier saturation.  
To address this issue, DFT-s-OFDM, which incorporates a DFT precoding stage, has been adopted for UL transmission since 4G due to its inherently lower PAPR. 
In the first release of 5G NR (Rel. 15), $\pi/2$-rotated
binary phase-shift keying ($\pi/2$-BPSK) modulation was introduced for UL DFT-s-OFDM transmission, together with frequency domain spectrum shaping (FDSS), implicitly  supported through relaxed spectral mask requirements. 
FDSS was chosen to be implemented  as a proprietary  transmitter-side operation, as only limited performance gains were observed with filter knowledge at the receiver~\cite{R1-1705060}. 
 The support of FDSS was later  extended  to quadrature
phase-shift keying (QPSK) in 5G-Advanced (Rel. 18), again indirectly via power boosting capabilities.  This trend toward reducing PAPR of 3GPP systems is expected to continue in 6G, driven by the need for improved coverage and energy-efficiency for both TN and NTN scenarios, as  reflected by recent  proposals within 3GPP 6G study for  extending DFT-s-OFDM to DL transmissions~\cite{R1-2506550}.

Despite the potential for subsequent PAPR reduction, 
$\pi/2$-BPSK with FDSS has not been widely adopted in practical deployments~\cite{R1-2506218}.  
One contributing factor is that pilot reference signals initially exhibited higher PAPR,
limiting the achievable power amplifier efficiency, although low-PAPR pilots were later introduced in 5G Rel. 16.  
In addition, as FDSS is a transparent transmitter-side technique,  
 no specific receiver assumption can be made,  which may lead to conservative shaping choices at the transmitter,  and  so not exploiting the full PAPR reduction potential~\cite{R1-2505913}. 

Furthermore, the use of $\pi/2$-BPSK  is constrained by modulation and coding scheme (MCS) tables in 5G NR~\cite{3GPP_TS_38214}.
In Release 15, $\pi/2$-BPSK is limited to low spectral efficiency regimes, with only a subset of MCS indices below 1 bit per channel use (bpcu) supporting this modulation\footnote{There are two tables for uplink data with DFT-s-OFDM~\cite{3GPP_TS_38214}. The first table has 13 MCS indices  below 1 bpcu  spectral efficiency, i.e. nearly half of the unreserved MCS, but only the 6 first  MSCs 
for up to 0.19 bpcu can be configured with  either $\pi/2$-BPSK or QPSK, while the 7 others with higher spectral efficiencies are only with QPSK. In the other table, $\pi/2$-BPSK is limited to the two first MCSs at 0.23 and 0.30 bpcu, while the 5 other MSCs below 1 bpcu can only be used with QPSK.}, while higher spectral efficiencies rely exclusively on QPSK or high-order modulations.  
This limitation is being relaxed in 5G Rel. 20, extending $\pi/2$-BPSK to higher spectral efficiencies up to 0.877 bpcu~\cite{R1-260XXXX},  targeting low PAPR level that are difficult  to achieve with QPSK. In particular, alternative low-PAPR approaches for higher-order modulations, such as FDSS with spectrum extension (FDSS-SE), inherently trade  spectral efficiency for PAPR reduction~\cite{Nokia21,PitavalFDSS-SE}.

\subsection{Performance Analysis of DFT-s-OFDM with FDSS}
For MCSs operating at low spectral efficiencies and low signal-to-noise ratio (SNR), the PAPR reduction provided by FDSS typically comes at negligible cost in terms of raw link performance.  
However, as the spectral efficiency increases, the PAPR benefit of FDSS is progressively offset 
by a degradation in raw link performance.  In practice, to limit this impairment, the level of spectral shaping is typically constrained, as reflected in 3GPP specifications through spectrum equalization masks.

Typical receiver implementations of DFT-s-OFDM rely on OFDM demodulation with single-tap (scalar)  equalization, followed by DFT de-spreading. Single-tap equalization is generally preferred due to its low complexity.
However, similar to channel fading fluctuation, FDSS   breaks the  orthogonality among  transmitted DFT-s-OFDM pulses, and contributes to inter-carrier interference (ICI)  and/or noise amplification after equalization. 
In contrast to OFDM, where the choice of single-tap equalizer has no impact on performance~\cite{ErkaiPIMRC24}, in DFT-s-OFDM  the equalizer choice plays a significant role.  
Among commonly used equalizers --  match filter (MF), zero-forcing (ZF), and minimum mean square error (MMSE) -- MMSE is typically preferred in practice~\cite{Nokia21} due to its observed better performance.  
 
As a result, FDSS in DFT-s-OFDM inherently introduces a trade-off between PAPR and signal-to-interference-plus-noise ratio (SINR). 
Despite its practical relevance,
the fundamental  
performance analysis 
of DFT-s-OFDM schemes   
under practical receiver assumptions, such as single-tap equalization, remains incomplete.    
In~\cite{Nisar2007},  SINR expressions for DFT-s-OFDM with independent and identical distributed (i.i.d.)  QAM constellation were derived for MMSE and ZF equalizers;  related  analyses can also be found in~\cite{SanchezVTC09,SanchezTVT2011,SanchezTVT2013}, as well as in more recent developments~\cite{SahinComLet2021}.  
Such SINR expressions are not only useful for performance evaluation but also for deriving log-likelihood ratio (LLR) metrics in practical receivers    
(although performance are often robust to LLR approximations).    
Extensions to $\pi/2$-BPSK  are not covered in \cite{Nisar2007}, and practical implementations typically rely on heuristic adaptations (e.g., scaling factors).  
In~\cite{KimTVT18}, further theoretical insights into $\pi/2$-BPSK DFT-s-OFDM with FDSS were provided, showing that the SINR depends on the rotation angle for rotated BPSK constellations, unlike the case for rotated QAM~\cite{TENCON18}. However, this analysis assumes a specific constellation-derotating equalizer and is limited to frequency-flat additive white Gaussian noise (AWGN) channels. 
In~\cite{WLNyquistTCOM21}, the Nyquist criteria was generalized to DFT-s-OFDM with FDSS and rotated pulse amplitude modulation. Without spectrum truncation or extension, and in the presence of channel fading, the resulting  widely-linear ZF in~\cite{WLNyquistTCOM21} reduces to the conventional single-tap ZF. More recently, \cite{WLMMSETWC24} investigated complexity reduction methods for multi-user frequency-domain  widely-linear matrix  equalization by exploiting signal properties of DFT-s-OFDM with FDSS and 
rotated constellations,   
and thus involves significantly higher complexity equalization than typically considered in practical systems. 

\subsection{Alternative Low-PAPR Modulations}
Traditionally, satellite systems have relied  on proprietary modulation schemes, typically based on a single carrier waveforms.
Due to stringent link budget constraints, these
systems predominantly employ low-order modulations,  
such as ($\pi/2$-) BPSK,  \mbox{($\pi/4$-)} QPSK, offset-QPSK (OQPSK), Gaussian minimum-shift keying (GMSK), or differentially-encoded (DE-) BPSK and QPSK~\cite{MajeedTCOM97,SaeedSurvey20,PositionSurvey22}.  
OQPSK, also referred to as staggered QPSK (SQPSK)~\cite{BoydTCOM19},  is a variant of QPSK in which the in-phase (I) and quadratic (Q) components of the carrier waveform are shifted by half a symbol period. As a result, the I and Q branches never change simultaneously, eliminating abrupt $\pi$ phase transitions and limiting the maximum phase variation to  $\pi/2$. This property reduces the PAPR while maintaining the same spectral efficiency as QPSK. 

 Many low-PAPR modulations belong to the family of 
continuous phase modulations (CPM)~\cite{AulinTCOM81}.  
In CPM, the carrier phase is modulated continuously with memory, resulting in a constant-envelope waveforms. MSK is a particular form of continuous-phase frequency-shift keying which can be interpreted as closely related to OQPSK:  bits are alternately mapped on the I and Q components, with the Q component delayed by half a symbol period. However, unlike OQPSK which employs rectangular pulse shaping, MSK uses half-sinusoid pulses. GMSK~\cite{MurotaTCOM81}, widely used in 2G GSM systems, extends MSK by applying
Gaussian filtering prior to modulation.   
Similarly, shaped OQPSK, filtered OQPSK, and Feher-patented QPSK~\cite{FeherTVT07}, are hybrid schemes   
that smooth phase transitions through
pulse shaping, achieving near-constant envelope signals.
DE modulations are 
a form of phase modulation  
where typically each constellation symbol depends of the previous one, enabling non-coherent detection. 

The low-PAPR single-carrier modulations  used in traditional satellite communications  are not directly compatible with OFDM-based waveforms.  
Attempts to incorporate CPM into OFDM, such as~\cite{ThompsonTCOM08}, typically require significant modifications to the transceiver architecture and increased receiver complexity.  
Similarly, the so-called OFDM/OQAM scheme~\cite{Bölcskei2003}
is  in fact, and despite its name, a filter-bank multicarrier (FBMC) system employing an advanced prototype filter, and therefore differs fundamentally from
conventional CP-OFDM as defined in 3GPP~\cite{BoydTCOM19}. 
Finally,  
coding-based approaches for PAPR reduction have
also been proposed, 
where the channel code is designed to map onto specific symbol codewords.  
However, such techniques often come at the cost of increased complexity or degraded error-rate performance~\cite{RahmatallahSurvey13}. 
Moreover, they are typically tailored to OFDM, optimized for specific configurations, and are not always scalable, while the achieved PAPR reduction is generally modest compared to DFT-s-OFDM.

\subsection{Contributions}
This article makes the following contributions.

First, we provide a theoretical analysis of DFT-s-OFDM with single-tap equalization, as commonly used in 5G systems.  
We derive general SINR expressions for DFT-s-OFDM with FDSS, applicable to arbitrary i.i.d. complex constellation, as well as to BPSK and $\pi/2$-BSPK.   
The results hold for general block-fading channels and arbitrary single-tap equalizers, thereby extending prior work in~\cite{Nisar2007} for QAM with MMSE and ZF equalization, and~\cite{KimTVT18} which considered rotated-BPSK with FDSS but 
in a frequency-flat AWGN channel.  
The derived SINR expressions enable theoretical bit error rate (BER) and mutual information analysis 
of DFT-s-OFDM with FDSS under practical receiver models.

Second, we propose 
a correlation-based spectral-shaping scheme,  termed repeated-and-offset QPSK (RO-QPSK), which relies solely on pre-modulation bit processing  
and is fully compatible with legacy DFT-s-OFDM transceivers.  
Specifically, the input bits  are Manchester encoded (repeated and flipped) with an offset between the I and Q branches prior to   
conventional QPSK mapping.  
This results in an order-one OQPSK-like modulation adapted to the DFT-s-OFDM structure. 
At the receiver, the induced redundancy is exploited through low-complexity symbol combining after DFT despreading and single-tap equalization.  

We show that this correlation structure of RO-QPSK induces a spectral shaping effect, leading to a Hann-window-shaped transmit spectrum without explicit filtering. 
This results in significantly reduced envelope fluctuations and a low PAPR on the order of 2 dB, comparable to $\pi/2$-BPSK with aggressive FDSS.
 
In addition, we derive the corresponding end-to-end SINR expressions for RO-QPSK, including the case where it is combined with FDSS, enabling a unified theoretical comparison with conventional $\pi/2$-BPSK and QPSK schemes.

Finally, we provide extensive link-level  
evaluations  across a wide range of spectral efficiencies, block lengths, channel conditions, and bandwidths.  
The impact of channel estimation errors and large residual frequency offsets beyond typical operational regimes is also investigated.   
The results show that RO-QPSK achieves performance gains over 5G-like $\pi/2$-BPSK  with FDSS at comparable PAPR levels in narrowband and weakly frequency-selective channels, as commonly encountered in satellite communications. In these scenarios, RO-QPSK approaches the performance of plain $\pi/2$-BPSK  (without FDSS) while maintaining low PAPR.  
In more frequency-selective conditions, the performance remains on par with existing schemes.

The remainder of this article is organized as follows. Section II describes the system model and presents results for DFT-s-OFDM with legacy constellations.  Section III introduces the RO-QPSK modulation and demodulation principles  and analyzes its key properties. Section IV provides a performance analysis in terms of PAPR, out-of-band (OOB) emission, uncoded bit error rate (BER), and mutual information. Section V presents block error rate (BLER) simulation results and evaluates the potential of RO-QPSK for satellite access. Section VI concludes the paper.

\section{System Model, Background and Preliminaries}

\subsection{DFT-s-OFDM Transmission}

\subsubsection{DFT precoding}
One DFT-s-OFDM symbol with $\Nsc$ DFT-s-OFDM subcarriers\footnote{DFT-s-OFDM is a multi-carrier system whose \emph{subcarriers} are time-multiplexed pulses that  are (possibly shaped) time-shifted Dirichlet kernels. In practice, DFT-s-OFDM is typically implemented as DFT-precoded OFDM, where OFDM itself is a multi-carrier system whose \emph{subcarriers} are frequency-multiplexed sinusoids. We will use  the term \emph{subcarrier} interchangeably for both systems and refer to \emph{ICI}   
accordingly, depending of the context.}  
is first computed by applying the DFT precoding of $\Nsc$ constellation symbols   as  
\begin{equation} \label{eq:X[k]}
X_k=\frac{1}{\sqrt{\Nsc}} \sum_{m=0}^{\Nsc-1}x[m]  
e^{-\jrm \frac{2\pi}{\Nsc}  km}	. 
\end{equation}
where $\jrm = \sqrt{-1}$ is the imaginary unit. 
The constellation symbols $\{x[m]\}$  are assumed to be zero-mean with unit average energy, i.e., $\expect{|x[m]|^2}=1$.
If the constellation symbols $\{x[m]\}$ are i.i.d.  
then the subcarriers coefficients $\{X_k\}$ are also i.i.d. with mean power $\expect{|X_k|^2 }=1$.

\subsubsection{OFDM with FDSS}
One CP-OFDM symbol is then computed using an inverse fast Fourier transform (IFFT) size of $\Nfft$ and a cyclic prefix (CP) of length  $\Ncp$ samples as 
\begin{equation} \label{eq:s[n]}
s[n]= \frac{\eta}{\sqrt{\Nfft}} \sum_{k=0}^{\Nsc-1}F_k X_k, 
e^{\jrm \frac{2\pi}{\Nfft}nk} 
\end{equation}
 $-\Ncp \leq n\leq \Nfft-1$, where $\Nsc$ is the number of modulated subcarriers among the $\Nfft$ inputs, $X_k$ are the data-dependent subcarrier coefficients, and  
$F_k$ is an FDSS window with average power $\mu_{F^2} = \frac{1}{\Nsc} \sum_{k=0}^{\Nsc-1} |F_k|^2$.

The variable $\eta$ is a power normalization factor ensuring $\expect{|s[n]|^2} = \Nsc/\Nfft$. In the case of i.i.d.  constellation symbols  $\expect{|s[n]|^2} = (\eta^2 \Nsc)/(\mu_{F^2}\Nfft)$, thus $\eta =  1/\sqrt{\mu_{F^2}}$, which simplifies to $\eta = 1$ when FDSS is not applied.

For evaluation, we use the deformed Hann window with  shaping parameter $0\leq \beta\leq 1$
\begin{equation} \label{eq:ModHanWin}
F_k =\frac{1}{\omega}\left(1-\frac{1-\beta}{1+\beta} \cos  \left(\frac{2\pi k+\pi}{\Nsc}\right)\right)   
\end{equation}
which we refer to as ``FDSS($\beta$)'' for convenience. Normalization with $\omega = \sqrt{1+\frac{(1-\beta)^2}{2(1+\beta)^2}}$ ensures $\mu_{F^2}=1$, and so when symbols are in addition i.i.d., then $\eta = 1$. This window corresponds to commonly used designs in 3GPP, where most aggressive power ripple considered is $\beta =-14$ [dB]~\cite{PitavalFDSS-SE}.

\subsection{Legacy QPSK-based Constellations} 
 We focus on QPSK-based constellations as illustrated in Table~\ref{tab:QPSK}.  
A bit $b_i$  modulates  either the I or Q component of a QPSK symbol  as
\begin{equation}
\alpha_i= \frac{1}{\sqrt{2}} (1-2 b_i ).
\end{equation}
Namely, $b_i=0$ maps to the amplitude $\alpha_i=\frac{1}{\sqrt{2}}$ while $b_i=1$ maps to the opposite direction $\alpha_i=-\frac{1}{\sqrt2}$. 

\subsubsection{Gray-mapped QPSK}
With Gray-mapped QPSK, each constellation symbol carries 2 bits as $x[m] = c^{\rm QPSK}(b_{2m},b_{2m+1})$ where 
\begin{equation}
c^{\rm QPSK}(b_i,b_{i+1})= \alpha_i+\jrm \alpha_{i+1}.
\end{equation}
The modulation is a direct serial-to-parallel conversion onto the I/Q branches.

\begin{table}[t] \centering
	\caption{Illustration of legacy 5G QPSK-based constellation}
	\label{tab:QPSK}
	\begin{tabular}{|c|c||c|c|}
	 \hline 
	\multicolumn{4}{|c|}{QPSK} \\\hline\hline
	\multicolumn{2}{|c||}{Bits} &\multicolumn{2}{c|}{Constellation symbols} \\
I-branch & Q-branch & I-branch & Q-branch\\ \hline
\cellcolor{blue!30} $b_0$ & \cellcolor{red!30}$b_1$ & \cellcolor{blue!30} $\alpha_0$ &\cellcolor{red!30}$\alpha_1$\\ 
\cellcolor{teal!30}  $b_2$ & \cellcolor{yellow!30} $b_3$ & \cellcolor{teal!30} $\alpha_2$ &$\cellcolor{yellow!30} \alpha_3$\\ 
\ldots & \ldots & \ldots & \ldots\\ 
\cellcolor{orange!40} $b_{2(\Nsc-1)}$ & \cellcolor{green!30}  $b_{2\Nsc-1}$ & \cellcolor{orange!40}$\alpha_{2(\Nsc-1)}$ &\cellcolor{green!40} $\alpha_{2\Nsc-1}$\\\hline 
\end{tabular}

\vspace{0.4cm}

	\begin{tabular}{|c|c||c|c|}
	 \hline
	\multicolumn{4}{|c|}{BPSK} \\\hline\hline
	\multicolumn{2}{|c||}{Bits} &\multicolumn{2}{c|}{Constellation symbols} \\
I-branch & Q-branch & I-branch & Q-branch\\  \hline
\cellcolor{blue!30}$b_0$ &\cellcolor{blue!30} $b_0$ & \cellcolor{blue!30}$\alpha_0$ &\cellcolor{blue!30}$\alpha_0$\\ 
\cellcolor{red!30}$b_1$ & \cellcolor{red!30}$b_1$ &$\cellcolor{red!30}\alpha_1$ &$\cellcolor{red!30}\alpha_1$\\ 
\ldots & \ldots & \ldots & \ldots\\ 
\cellcolor{olive!30}$b_{\Nsc-1}$ &\cellcolor{olive!30} $b_{\Nsc-1}$ &\cellcolor{olive!30} $\alpha_{\Nsc-1}$ &\cellcolor{olive!30}$\alpha_{\Nsc-1}$\\\hline 
\end{tabular}

\vspace{0.4cm}

	\begin{tabular}{|c|c||c|c|}
	 \hline
	\multicolumn{4}{|c|}{$\pi/2$-BPSK} \\\hline\hline
	\multicolumn{2}{|c||}{Bits} &\multicolumn{2}{c|}{Constellation symbols} \\
I-branch & Q-branch & I-branch & Q-branch\\ \hline
\cellcolor{blue!30} $b_0$ & \cellcolor{blue!30} $b_0$ &\cellcolor{blue!30} $\alpha_0$ & \cellcolor{blue!30}$\alpha_0$\\ 
\cellcolor{red!30}  \color{lightgray!20}$\overline{b_1}$ & \cellcolor{red!30}$b_1$ & \cellcolor{red!30} \color{lightgray!20} $-\alpha_1$ & \cellcolor{red!30}$\alpha_1$\\ 
\cellcolor{teal!30}  $b_2$ & \cellcolor{teal!30}$b_2$ & \cellcolor{teal!30}$\alpha_2$ & \cellcolor{teal!30} $\alpha_2$\\
\ldots & \ldots & \ldots & \ldots\\ 
\cellcolor{cyan!30} $b_{\Nsc-2}$ &\cellcolor{cyan!30} $b_{\Nsc-2}$ & \cellcolor{cyan!30} $\alpha_{\Nsc-2}$ & \cellcolor{cyan!30}$\alpha_{\Nsc-2}$\\ 
\cellcolor{olive!30}\color{lightgray!20} $\overline{b_{\Nsc-1}}$ & \cellcolor{olive!30}$b_{\Nsc-1}$ &\cellcolor{olive!30} \color{lightgray!20}$-\alpha_{\Nsc-1}$ & \cellcolor{olive!30}$\alpha_{\Nsc-1}$\\\hline 
\end{tabular}
\end{table}

\subsubsection{BPSK }
With BPSK, each constellation symbol carries only 1 bit as $x[m] = c^{\rm BPSK} \left(b_m\right)$ where
\begin{eqnarray}
c^{\rm BPSK} \left(b_m\right)&=&  \alpha_m (1+\jrm) = \sqrt{2} \alpha_m e^{\jrm \frac{\pi}{4}} \nonumber \\ 
 &=& c^{\rm QPSK}(b_m,b_m). 
\end{eqnarray}
BPSK can thus be regarded as  QPSK where bits are encoded with half-rate repetition coding prior to serial-to-parallel conversion, as illustrated in Table~\ref{tab:QPSK}.

\subsubsection{$\pi/2$-BPSK}
With  $\pi/2$-BPSK, the transmitted symbols,  $x[m] = c^{\pi/2 {\rm -BPSK} }(b_m)$, are BPSK symbols with a $\pi/2$-rotation along the  subcarriers as\footnote{The definition of $\pi/2$-BPSK here follows the 3GPP NR standard. Alternative definitions can be found in the literature, such as $c^{\pi/2 {\rm -BPSK} }(b_m) =
e^{\jrm \frac{\pi}{2} m} c^{\rm BPSK}(b_m)$~\cite{KimTVT18}, which differs only by certain sign changes that do not affect the signal characteristics.}  
\begin{eqnarray} 
c^{\pi/2 {\rm -BPSK} }(b_m) &=&
e^{\jrm \frac{\pi}{2} (m \ {\rm mod}\ 2)} c^{\rm BPSK}(b_m) \nonumber\\
&=&
\begin{cases} 
 c^{\rm QPSK}(b_m,b_m)&  i \text{ even} \\
 c^{\rm QPSK}  (\overline{b_m},b_m)& m \text{ odd} 
\end{cases}
\end{eqnarray}
where $\overline{b_m}= 1\oplus b_m $ is the binary complement. 
$\pi/2$-BPSK can therefore be interpreted as QPSK with half-rate coding with bit-flip repetition as shown in Table~\ref{tab:QPSK} (assuming $\Nsc$ is even).

\subsection{DFT-s-OFDM Demodulation}

\subsubsection{Channel}
The signal is received through a multi-path channel with $L_p < \Ncp$ taps for indices 
$0\leq n \leq \Nfft-1$ as
\begin{equation} 
y[n]=\sqrt{\snr} \sum_{l=0}^{L_p-1} h_l [n]s[n-l]  +z[n].
\end{equation}
The time-domain noise $z[n]$ is assumed standard circularly symmetric complex Gaussian with zero-mean and unit average energy, i.e., $\expect{|z[n]|^2}=1$.  The average  channel energy is assumed to satisfy $\sum_{l=0}^{L_p-1} \expect{ |h_l[n]|^2 } = 1$.

\subsubsection{OFDM Demodulation} 
The received signal is first demodulated by an FFT after discarding the CP. In general, with a time-varying channel, the demodulated OFDM symbol for the $k$-th subcarrier can be written as~\cite{ErkaiPIMRC24}
\begin{equation} \label{eq:Yk}
Y_{k} = \eta\sqrt{\snr} F_{k} H_{k,k} X_k + I_k  + Z_k
\end{equation}
where the OFDM ICI term is 
\begin{equation} 
I_k = \sum_{k'= 0, k' \neq k}^{\Nsc-1}\eta \sqrt{\snr}  F_{k'}  H_{k,k'} X_{k'}
\end{equation}
 and the channel components $H_{k, k'}$ are the $(k-k')$-th DFT coefficients
\begin{equation}
    H_{k,k'} = \frac{1}{\Nfft} \sum_{n = 0}^{\Nfft-1} H_{k'}[n] e^{j 2\pi \frac{n(k - k')}{\Nfft}}
\end{equation}
of the time-varying  discrete Fourier response of the channel impulse response (CIR), 
\begin{equation}
    H_m[n] = \sum_{l=0}^{L_p-1} h_l[n] e^{-j 2\pi \frac{m l}{\Nfft}},
\end{equation}
and  $Z_k=  \frac{1}{\sqrt{\Nfft}} \sum_{n=0}^{\Nfft-1} z[n] e^{j 2\pi \frac{-k n}{\Nfft}}$ is the frequency-domain  noise with unit variance, i.e., $\expect{|Z_k|^2}=1$.

OFDM has been designed under the assumption of a time-invariant channel within one OFDM symbol. However, if the channel  varies, the terms $  H_{k,k'}$ with $k\neq k'$ are non-zero and correspond to OFDM ICI. 
In practice, channel variation within one OFDM symbol is typically small, even under moderate Doppler effect, and the OFDM ICI terms can generally be considered negligible ($|H_{k, k' \neq k} |\approx 0$) or modeled as part of the noise.

We  will consider the conventional approach of one-tap equalization, based on the assumption that the channel taps remain constant over each OFDM symbol, i.e. $h_l\left[n\right]=h_l$, and that the CP is longer than the maximum delay spread.  
 Under these assumptions, the received symbol on subcarrier $k$ after FFT demodulation can be written as
\begin{equation}
Y_k=\widetilde{H}_k X_k+Z_k
\end{equation}
with the effective subcarrier channel coefficient
\begin{equation}
\widetilde{H}_k= \eta \sqrt{\snr} F_k H_k
\end{equation} 
where $H_k=\sum_{l=0}^{L_p-1}h_l e^{-\jrm 2\pi\frac{kl}{\Nfft}}$ is the (unnormalized) DFT of the CIR at subcarrier $k$.

\subsubsection{One-tap Equalization} 
The received subcarrier coefficients are then equalized as 
\begin{eqnarray}
\widetilde{Y}_k &=& E_k Y_k \nonumber\\
&=& G_k X_k+E_k Z_k.
\end{eqnarray}
such that the equalized channel gain is real-valued\footnote{With channel estimation errors, equalization is such that  the \emph{estimated} equalized channel is real. The true equalized channel may therefore not be strictly real, leading to some mismatch and performance degradation. However, equalization and LLR computation can only rely on the available channel estimate.},
\begin{equation}
G_k= E_k\widetilde{H}_k \in \mathbb{R} \label{eq:RealEq} . 
\end{equation}
For convenience, we extend the indexing notation modulo $\Nsc$, such $G_{h} = G_k$ if $h = k \pmod \Nsc$.


  
Typical equalizers satisfying~\eqref{eq:RealEq}  include 
MF, ZF, and MMSE, 
defined respectively as
\begin{equation}
    E_k = 
    \begin{cases}
        \widetilde{H}_k^{*},   &\text{MF} \\  
			\widetilde{H}_k^{-1}, &\text{ZF} \\
        \frac{ \widetilde{H}_k^*}{| \widetilde{H}_k |^2 + 1}, &\text{MMSE}   
						\end{cases}.
\end{equation}

\subsubsection{DFT De-spreading}

After equalization, the inverse DFT precoding is applied on the equalized subcarrier coefficients $\widetilde{Y}_k$, yielding the received symbols 
\begin{eqnarray}
r[m]&=& \frac{1}{\sqrt{\Nsc}} 
\sum_{k=0}^{\Nsc-1}\widetilde{Y}_k e^{\jrm\frac{2\pi}{\Nsc}km} \nonumber\\
&=& \frac{1}{\sqrt{\Nsc}} \sum_{k=0}^{\Nsc-1} G_k X_k 
e^{\jrm\frac{2\pi}{\Nsc}km} +n[m] 
\end{eqnarray}
with post-processed noise term
\begin{equation} \label{eq:n[m]}
n[m]=\frac{1}{\sqrt{\Nsc}}\sum_{k=0}^{\Nsc-1}E_kZ_ke^{\jrm \frac{2\pi}{\Nsc}km}.	
\end{equation} 
Replacing  $ X_k$ by~\eqref{eq:X[k]} we obtain
\begin{equation} \label{eq:r[m]Mat}
r[m] =\sum_{n=0}^{\Nsc-1}x[n] g_{m-n}  + n[m]
\end{equation}
where
\begin{equation} \label{eq:gm}
g_{m}=\frac{1}{\Nsc} \sum_{k=0}^{\Nsc-1} G_ke^{\jrm \frac{2\pi}{\Nsc} km}.
\end{equation}

\subsection{Effective SINR With i.i.d. Complex Symbols}
We consider a typical receiving scheme where symbols are demodulated independently on each subcarrier, and the contributions from other subcarriers are treated as interference.
Isolating the desired symbol on a given subcarrier, 
the received symbols can be written as
\begin{equation}\label{eq:r[m]}
r[m]=\mu_{G}  x[m]+\ICI[m]+ n[m]
\end{equation}
where the useful channel component 
\begin{equation}\mu_{G} \triangleq  g_{0} = 
  \frac{1}{\Nsc} \sum_{k=0}^{\Nsc-1}  G_k \label{eq:mu(EH)}
\end{equation}
 is independent of $m$ and 
DFT-s-OFDM ICI resulting from the OFDM equalization is
\begin{equation} \label{eq:ICI[m]}
\ICI[m] = \sum_{\substack{n=0 \\ n\neq m}}^{\Nsc-1} g_{m-n}x[n]. 
\end{equation}

We have the following result proved in Appendix A-A. 
\begin{Prop} \label{lem:SINR}
The effective SINR of~\eqref{eq:r[m]} with i.i.d. complex symbols  is 
\begin{equation}
\SINR^{\rm{iid}\mathbb{C}} = \frac{\mu^2_{G}}{\var_{G} + \mu_{ |E|^2}},
\end{equation}
where the noise power is 
\begin{equation}
\mu_{|E|^2}=\frac{1}{\Nsc} \sum_{k=0}^{\Nsc-1} |E_k|^2   \label{eq:muE}
\end{equation} and the interference power is 
\begin{equation} 
\var_{G} = \frac{1}{\Nsc} \sum_{k=0}^{\Nsc-1}  G^2_k -\mu^2_{G}  \label{eq:varG} .
\end{equation}
\end{Prop}

As indicated by the notations, the signal power and interference power correspond, respectively, to the mean and variance of the equalized subcarriers channel gains $\{G_k\}$. DFT-s-OFDM ICI results from the residual FD channel variation after OFDM equalization.  The noise power reflects the average squared magnitude of the equalizer coefficients.

Prop.~\ref{lem:SINR} generalizes the results in~\cite{Nisar2007}. For ZF and MMSE, we recover the following (see Appendix A-B.) 
\begin{Cor}[\cite{Nisar2007}] \label{Cor:ZF_MMSE}
  With ZF, the SINR is 
\begin{equation}
\SINR^{\rm{iid}\mathbb{C}}_{\rm ZF} =\frac{\Nsc}{\sum_{k=0}^{\Nsc-1} |\widetilde{H}_k|^{-2}},
\end{equation}
and with MMSE
\begin{equation} \label{eq:SINR_MMSE}
\SINR^{\rm{iid}\mathbb{C}}_{\rm MMSE} =\frac{\mu_{G}}{1-\mu_{G}}. 
\end{equation}
\end{Cor}

Contrary to OFDM~\cite{ErkaiPIMRC24}, the choice of equalizer significantly impacts the performance of DFT-s-OFDM. 
As is well known, when $\snr \to 0$, MMSE $\to$ MF, while when $\snr \to \infty$, then MMSE $\to$ ZF.  
Here, we are interested in low-order modulation operating in low SNR, for which 
in general, MMSE provides the best performance.

When considering Gray-mapped QPSK transmission, soft-symbol demodulation can be performed independently on the I and Q branches as
\begin{equation} 
\mu_{G} \alpha_m \approx
\begin{cases} 
\Re \left\{ r[m] \right\} &  m \text{ even} \\
 \Im \left\{r[m]\right\}&   m \text{ odd} 
\end{cases}.
\end{equation}

\subsection{Effective SINR With $\pi/2$-BPSK}

When transmitting BPSK or $\pi/2$-BPSK symbols, the effective SINR can be improved as the symbol power is fully allocated on a single dimension, and thus half of the noise power from the other dimension can be filtered out. The effective interference, being also filtered, is  slightly modified. 

If BPSK is transmitted, the received signal $r[m]$ in~\eqref{eq:r[m]} is  rotated back to the I-branch, and the Q-branch is filtered out as $r_{\rm R}[m]= \Re \left\{ e^{-\jrm \frac{\pi}{4}}  r[m] \right\}$. For $\pi/2$-BPSK, the $\pi/2$-precoding of the constellation must in addition be inverted, as 
$r_{\rm R}[m]= \Re\left\{e^{-\jrm \frac{\pi}{2}\left(\frac{1}{2}+ (m  \ {\rm mod}\ 2)\right)} r[m]\right\}$. 
We then obtain 
\begin{equation} \label{eq:rR[m]}
r_{\rm R}[m]= \mu_{G} \sqrt{2}\alpha_m +  \ICI_{\rm R}[m] +   n_{\rm R}[m] 
\end{equation}
where the ICI term is real and given by  
$ \ICI_{\rm R}[m] =  \sum_{\substack{n=0 \\ n\neq m}}^{\Nsc-1}  \sqrt{2}\alpha_n  g_{m-n}^{\rm R} $ with    
$g_{m-n}^{\rm R} = \Re\left\{ g_{m-n}\right\}$ or $\Re\left\{e^{-\jrm \frac{\pi}{2}\left(m  \ {\rm mod}\ 2\right)}  g_{m-n}\right\} $, and the noise is $ n_{\rm R}[m]= \Re \left\{ e^{-\jrm \frac{\pi}{4}}  n[m]\right\}  $ or $ \Re\left\{e^{-\jrm \frac{\pi}{2}\left(\frac{1}{2}+ (m  \ {\rm mod}\ 2)\right)}  n[m]\right\}  $ for BPSK and $\pi/2$-BPSK, respectively. 
We have the following result, proved in Appendix A-C. 
\begin{Prop} \label{lem:SINR_BPSK}
If $\Nsc$ is even,  the effective SINR of \eqref{eq:rR[m]} with i.i.d. ($\pi/2$-)BPSK symbols is 
\begin{equation}
\SINR^{\rm BPSK} = \frac{\mu^2_{G}}{\zeta^2_{G} + \frac{1}{2}\mu_{ |E|^2}},
\end{equation}
where the interference power is 
\begin{equation} \label{eq:zeta}
\zeta^2_{G}  = \begin{cases}\displaystyle 
\frac{1}{2\Nsc} \sum_{k=0}^{\Nsc-1}  G_k( G_{\Nsc-k}  +G_k)  -  \mu^2_{G}  & (\text{BPSK}) \\
\displaystyle  \frac{1}{2\Nsc} \sum_{k=0}^{\Nsc-1}  G_k( G_{\frac{\Nsc}{2}-k}  +G_k)  -  \mu^2_{G} & (\pi/2\text{-BPSK}) 
\end{cases} . 
\end{equation}
\end{Prop}

The ICI power for BPSK is related to that of  QPSK as $ \zeta^2_{G}  =\sigma_G^2 + \frac{1}{2\Nsc} \sum_{k=0}^{\Nsc-1}  G_k( G_{\Nsc-k}  -G_k)$. 
If the channel spectrum is approximately constant and the FDSS approximately symmetric, as is the case of most conventional window designs, then $G_{\Nsc-k} \approx G_k$, $ \zeta^2_{G} \approx \sigma_G^2$,  and $\SINR^{\rm BPSK} \approx \frac{\mu^2_{G}}{\sigma^2_{G} + \frac{1}{2}\mu_{ |E|^2}}$, i
.e., the difference from QPSK lies in a halved noise power.   

A noteworthy property of $\pi/2$-BPSK is that its performance with DFT-s-OFDM differs from BPSK; this seems to be little known  even though  though some earlier works  such as~\cite{KimTVT18} considered the  SINR impact of a phase-rotation over BPSK.  
For $\pi/2$-BPSK,  the ICI power relative to QPSK becomes $ \zeta^2_{G}  =\sigma_G^2 + \frac{1}{2\Nsc} \sum_{k=0}^{\Nsc-1}  G_k(G_{\frac{\Nsc}{2}-k}  -G_k)$. For  a flat channel without FDSS, one again has  $ \zeta^2_{G} \approx  \sigma_G^2$; however, this no longer holds with symmetric FDSS.  Evaluations show that,  with commonly used FDSS windows, $\pi/2$-BPSK outperforms BPSK, and compared to QPSK, it not only halve the noise power, but also mitigate the ICI power, i.e., $ \zeta^2_{G}  <\sigma_G^2 $.  
This effect is specific to rotated-BPSK; otherwise constellation rotation has no impact on the SINR,  as for example observed in~\cite{TENCON18} for rotated QPSK. 

For both BPSK  and $\pi/2$-BPSK, in the low-SNR regime of interest where the ICI power is small compared to the noise power $\SINR^{\rm BPSK} \approx 2 \SINR^{\rm{iid}\mathbb{C}}$,  thus providing approximately twice the SINR, albeit at the cost of a halved data rate compared to QPSK. This leads to approximately the same Shannon capacity in the low-SNR regime, since $\frac{1}{2} \log_2(1+ \SINR^{\rm BPSK}) \approx \frac{1}{\ln 2} \SINR^{\rm{iid}\mathbb{C}} \approx \log_2(1+ \SINR^{\rm{iid}\mathbb{C}})$ as $\snr \to 0 $.

\section{Repeated-and-Offset QPSK for DFT-s-OFDM}

\subsection{Motivations} 
RO-QPSK is directly inspired by OQPSK used in single carrier waveform, but tailored for DFT-s-OFDM. 
\subsubsection{Limited Offset Possibility}
In conventional OQPSK, the I and Q branches carry independent bit streams with an offset of half a symbol period. This ensures that the phase shift of the combined signal never exceeds $\pi/2$, thereby reducing the PAPR compared to QPSK. However, an half-symbol offset can only be done in single carrier or FBMC systems, but not in conventional DFT-s-OFDM (implemented as   DFT-precoded OFDM). To circumvent this, we instead repeat the symbols and apply  a twice larger offset of one full symbol period, which effectively achieves a similar effect but reduces the data rate by half compared to QPSK -- analogous to $\pi/2-$BPSK.

\subsubsection{A Zero-Crossing Problem?}
With BPSK, consecutive symbols may have a phase shift of either 0 or $\pi$, while with QPSK, the phase shift between consecutive symbols can be 0, $\pi/2$, $\pi$, or $3\pi/2$. In the literature, these constellations are often described as suffering from zero crossings, i.e., a $\pi$-phase change may happen from one symbol to the next, resulting in high PAPR. Therefore, $\pi$/2-BPSK is often explained to achieve lower PAPR by avoiding zero-crossing:  consecutive symbol phases  systematically differ by $\pi/2$ instead. 
This interpretation is correct for single-carrier waveforms but is misleading for DFT-s-OFDM. DFT-s-OFDM subcarriers correspond to time-shifted sinc-like pulses, and consecutive subcarriers (or pulses) already exhibit a phase difference close to $\pi$~\cite{KimTVT18,TCOM25Pitaval}.  
Hence, taking into account this intrinsic subcarrier phase variation, a zero-crossing in DFT-s-OFDM actually occurs when two identical constellation symbols are transmitted consecutively. $\pi$/2-BPSK reduces the PAPR in DFT-s-OFDM not because it avoids $\pi$ phase changes, but because it avoids consecutive symbols with a 0 phase change. 
To obtain a similar effect with RO-QPSK, we introduce a sign alternation to ensure that consecutive symbols have a phase shift of either $\pi/2$ or $\pi$, but never 0, therefore maximizing the consecutive symbol phase variation.

\subsection{RO-QPSK Modulation}
Following these observations and assuming $\Nsc$ even, we consider a RO-QPSK modulation defined by mapping bits to QPSK symbols according to 
\begin{eqnarray} 
c^{\rm RO-QPSK }(b_{m-1}, b_m) &=&
\begin{cases} 
 c^{\rm QPSK}(b_m,\overline{b_{m-1}})&  m \text{ even} \\
 c^{\rm QPSK} (\overline{b_{m-1}},b_m) & m \text{ odd} 
\end{cases} 
\nonumber \\
 &=&
\begin{cases} 
\alpha_m -\jrm \alpha_{m-1}&  m \text{ even} \\
-\alpha_{m-1} +\jrm \alpha_{m} & m \text{ odd} 
\end{cases}.
\end{eqnarray}
For  clarity, indexing is defined modulo $\Nsc$  such that $b_{-1}=b_{\Nsc -1}$, and $\alpha_{-1}= \alpha_{\Nsc-1}$.
The transmitted symbol on the $m$-th DFT-s-OFDM subcarrier is then
\begin{eqnarray}
x[m] \!\!\! &=& \!\!\!c^{\rm RO-QPSK }(b_{m-1}, b_m) \nonumber \\ 
\!\!\! &=&\!\!\! (-1)^{m}\left(\alpha_{m-(m \text{ mod } 2)}- \jrm \alpha_{m-1+(m \text{ mod } 2 )}\right).	
\end{eqnarray} 
The resulting modulation is illustrated in Table~\ref{tab:RO-QPSK}.

\begin{table}[t] \centering
	\caption{  } 
	\label{tab:RO-QPSK}
\begin{tabular}{|c |c||c |c|}
\hline
	\multicolumn{4}{|c|}{RO-QPSK} \\\hline
	\multicolumn{2}{|c||}{Bits} &\multicolumn{2}{c|}{Constellation symbols} \\
I-branch & Q-branch & I-branch & Q-branch\\\hline
\cellcolor{blue!30} $b_0$ & \cellcolor{olive!30} \color{lightgray!20}$\overline{b_{\Nsc-1}}$ & \cellcolor{blue!30} $\alpha_0$ &\cellcolor{olive!30}\color{lightgray!20}  $ -\alpha_{\Nsc-1}$\\
\cellcolor{blue!30} \color{lightgray!20}$\overline{b_0}$ & \cellcolor{red!30} $b_1$& \cellcolor{blue!30}\color{lightgray!20} $-\alpha_0$ &\cellcolor{red!30}$\alpha_1$\\ 
\cellcolor{teal!30} $b_2$ & \cellcolor{red!30} \color{lightgray!20}$\overline{b_1}$ &\cellcolor{teal!30} $\alpha_2$ & \cellcolor{red!30}\color{lightgray!20} $-\alpha_1$\\
\cellcolor{teal!30} \color{lightgray!20}$\overline{b_2}$ & \ldots  &\cellcolor{teal!30} \color{lightgray!20}$-\alpha_2$ & \ldots\\
           \ldots & \ldots & \ldots & \ldots\\
 \ldots  &  \cellcolor{purple!30} $b_{\Nsc-3}$ &  \ldots  &\cellcolor{purple!30}$\alpha_{\Nsc-3}$\\
\cellcolor{cyan!30} $b_{\Nsc-2}$ &  \cellcolor{purple!30} \color{lightgray!20}$\overline{b_{\Nsc-3}}$ & \cellcolor{cyan!30} $\alpha_{\Nsc-2}$ &\cellcolor{purple!30}\color{lightgray!20}$-\alpha_{\Nsc-3}$\\
\cellcolor{cyan!30} \color{lightgray!20}$\overline{b_{\Nsc-2}}$ & \cellcolor{olive!30} $b_{\Nsc-1}$ & \cellcolor{cyan!30}\color{lightgray!20}$-\alpha_{\Nsc-2}$ & \cellcolor{olive!30} $\alpha_{\Nsc-1}$\\ \hline
\end{tabular}
\end{table}

While RO-QPSK is inspired from OQPSK, there are several important differences. 
In legacy single-carrier OQPSK, the transmitted waveform does not correspond to standard QPSK modulation, whereas in RO-QPSK the transmitted symbols remains standard QPSK symbols.
  
The bit-flip operation combined with symbol repetition introduces alternating signs in the I or Q components of successive symbols,
thereby compensating for the intrinsic near-$\pi$ phase changes between consecutive DFT-s-OFDM pulses, as explained previously.  

The joint use of  repetition and bit-flipping corresponds exactly a Manchester encoding applied to the bit streams of the I and Q branches. 
This establishes a direct connection to Manchester-encoded DFT-s-OFDM-based on-off keying~\cite{TCOM25Pitaval},  recently specified   in 5G-Advanced for low-power wake-up signals. 
The bit-to-symbol mapping that creates the I/Q offset can be interpreted as a specific form of interleaving.  

\begin{figure}
\vspace{-0.0cm} 
\centering
\includegraphics[width=.3\textwidth]{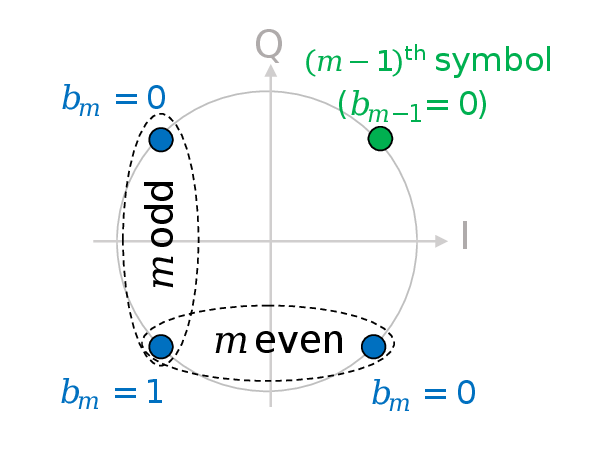}
\vspace{-0.2cm}
	\caption{Illustration of the differential encoding aspect of RO-QPSK. \label{fig:diffDiagram}}
\vspace{-0.3cm}
\end{figure}

Finally, RO-QPSK can also be interpreted as a form of DE-QPSK since each constellation symbol $x[m]$ depends on the previous symbol $x[m-1]$. An illustration is provided in Fig.~\ref{fig:diffDiagram}, showing how the selection of the current QPSK symbol depends on the preceding one.
Nevertheless, as will be shown,  symbols can still be decoded independently, avoiding the typical error propagation associated with differential modulations. 


\subsection{Spectrum and Power Normalization}

The following minor properties of RO-QPSK shall be observed:
\begin{Rem}
With RO-QPSK, the first subcarrier coefficient is always zero, i.e., $X[0]=0$.    
\end{Rem}
We do not exploit this property in this paper, but it could potentially be leveraged, for example, either i) as a side information to provide a small detection gain, ii) alternatively, to insert a reference symbol\footnote{A quick evaluation of this alternative indicated that it would increase the PAPR of RO-QPSK by approximately 1dB for a narrow band with $\Nsc = 24$, but only by about $0.3$dB in a wider band such as $\Nsc = 144$.}.

A  more notable characteristic, which differs significantly from the case of i.i.d. QAM,   is  that even without FDSS the power spectrum is not flat 
(see Appendix B-A.). 
\begin{Lem} \label{lem:PSD_RO-QPSK}
The power spectrum of RO-QPSK is characterized by $\expect{X_k X_h^*} = w_k \delta_{k-h} $ with 
Hann weight function 
\begin{equation}
w_k= 1-\cos{\frac{2\pi}{\Nsc}k}. 
\end{equation}
\end{Lem} 
Furthermore, the power normalization of RO-QPSK is 
(see Appendix B-B.).
\begin{Lem} \label{lem:PwNor_RO-QPSK}
The normalization factor for RO-QPSK is $\eta =\frac{1}{\sqrt{\mu_{w,F^2}}} $ where $\mu_{w,F^2} = \frac{1}{\Nsc} \sum_{k=0}^{\Nsc-1} w_k |F_k|^2$ is the Hann-weighted mean of the FDSS window. Without FDSS,  $\eta = 1$. 
\end{Lem} 

\begin{figure}[t] \centering
\vspace{-0.0cm} 
\includegraphics[width=.5\textwidth]{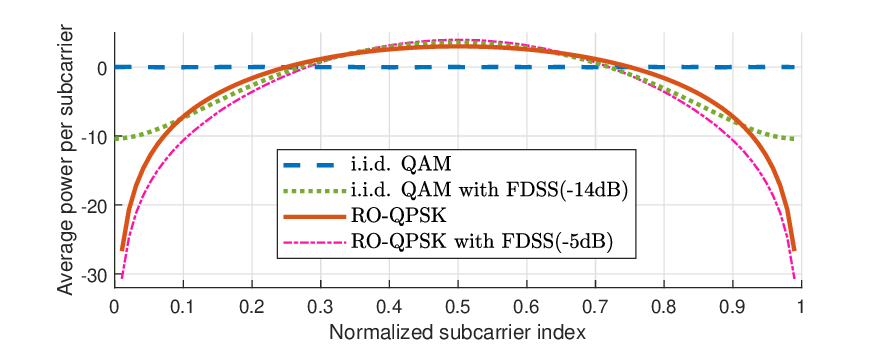}
\vspace{-0.4cm}
	\caption{Average subcarrier power for RO-QPSK and i.i.d. QAM, with and without FDSS.  \label{fig:AverageScPower}}
\vspace{-0.3cm}
\end{figure}

Fig.~\ref{fig:AverageScPower} shows the average  subcarrier power for i.i.d. QAMs (including $\pi$/2 BPSK) and RO-QPSK, with or without FDSS, as function of the normalized subcarrier index. The subcarrier coefficients are scaled according to their corresponding power normalization, i.e.,  the plot shows $\{ \eta^2 \expect{|X_k|^2} \}_{k=0}^{\Nsc-1} \}$ as function of $k/\Nsc$.  
Here  $\Nsc =96$ is used, but the overall shape remains consistent for other $\Nsc$, corresponding to different quantizations of the same underlying Hann function.

\subsection{RO-QPSK Demodulation}
The received symbols after DFT-s-OFDM demodulation are obtained as in~\eqref{eq:r[m]Mat} and leading to~\eqref{eq:r[m]}. 
With RO-QPSK, it can be observed that 
\begin{equation}
\mu_{G} \alpha_m \approx
\begin{cases}
\Re \left\{\frac{r[m]-r[m+1]}{2}\right\}& m \text{ even} \\
\Im \left\{\frac{r[m]-r[m+1]}{2}\right\}&m \text{ odd} 
\end{cases}, 
\end{equation}
so that consecutive symbols (modulo $\Nsc$)  can be combined to estimate the individual symbol components $\alpha_{m}$ on the I and Q branches.

Altogether, the post-combined demodulation can be expressed by taking only the even indices $m = 2l$, with $l=0,\ldots, \Nsc/2-1$,  as 
\begin{equation} \label{eq:rtilde[l]}
\tilde{r}[l]=  
\Re \left\{  \text{\scriptsize $ \frac{r[2l]-r[2l+1]}{2}$} \right\} +\jrm
 \Im \left\{  \text{\scriptsize $ \frac{r[2l+1]-r[2l+2]}{2}$}\right\}.  
\end{equation}
The resulting $\Nsc/2$ combined symbols in~\eqref{eq:rtilde[l]} can then be directly provided to a standard QPSK demodulator as if demodulating the set  $\{ q_l\}_{l=0}^{\frac{\Nsc}{2}-1}$ of equivalent QPSK symbols defined as 
\begin{eqnarray} \label{eq:RO-QPSKqpsksymb}
 q_l&=&  \alpha_{2l}+\jrm \alpha_{2l+1} \\ 
 &=& c^{\rm QPSK}(b_{2l},b_{{2l}+1}).
\end{eqnarray}

\begin{figure*}[t]
\vspace{-0.0cm} 
\centering
\includegraphics[width=0.8\textwidth]{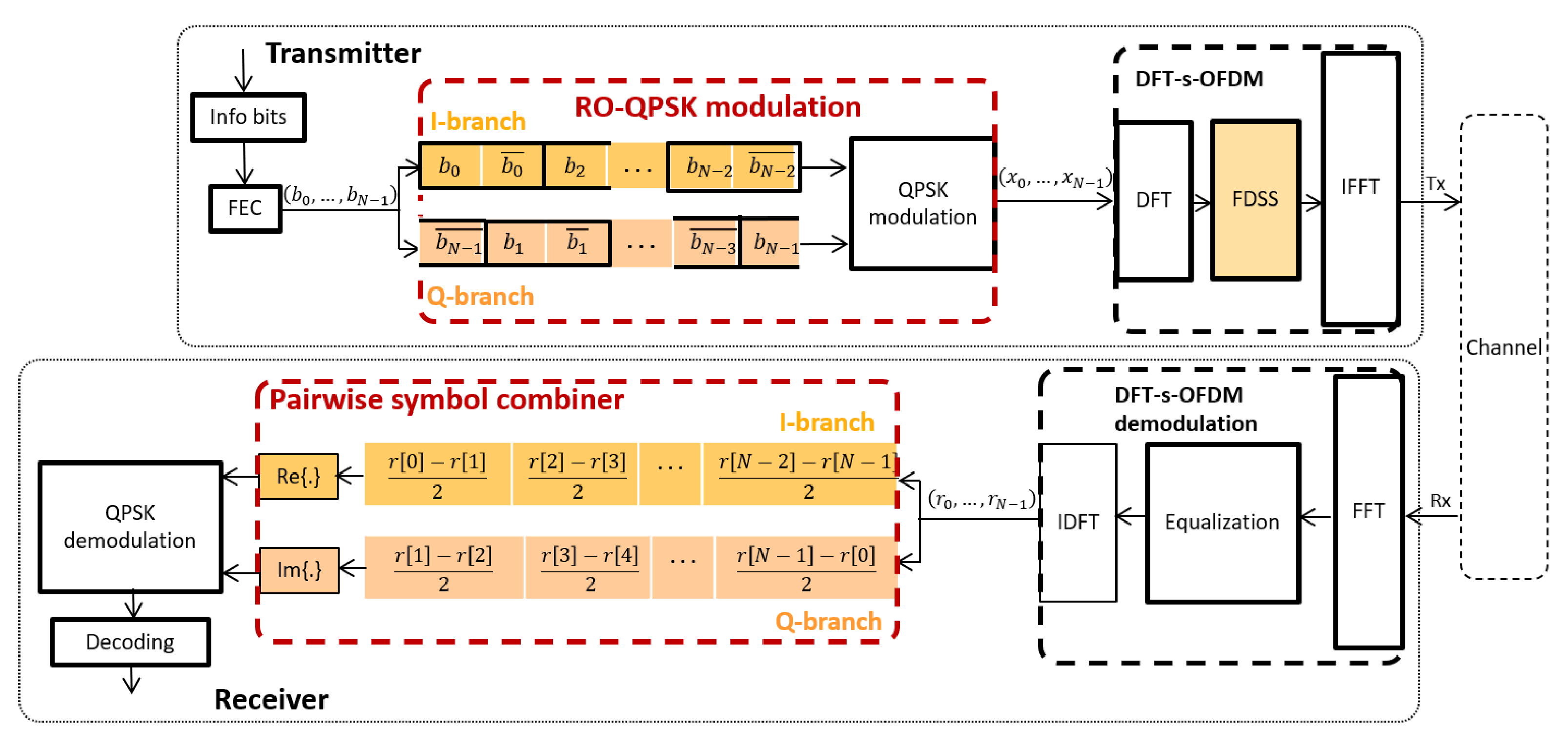}
\vspace{-0.2cm}
	\caption{RO-QPSK modulation and demodulation process.   \label{fig:TxDiagram}}
\vspace{-0.3cm}
\end{figure*}

Fig.~\ref{fig:TxDiagram}  summarizes the  transmitter-receiver pair of RO-QPSK. 
The additional implementation complexity is marginal at both the transmitter and the receiver. At the transmitter, no operations are required over the modulated complex symbols; only simple pre-modulation bit processing is performed. At the receiver, the combining can be applied separately on the I and Q branch. The scaling factor can be absorbed into the LLR computation, such that the additional processing reduces to  $\Nsc$ real additions. This is significantly lower than, for example FDSS or frequency-domain equalization, which typically require   $\Nsc$ complex multiplications.

\subsection{Effective SINR for RO-QPSK}
Expressing the pre-combined received symbols $r[m]$ as in~\eqref{eq:r[m]}, the post-combined received symbols~\eqref{eq:rtilde[l]} can be written as  
\begin{equation} \label{eq:rtilde1[l]}
\tilde{r}[l] = \mu_{G}q_l   +\widetilde{\ICI}_l+\widetilde{\rm n}_l 	
\end{equation}
where  
\begin{equation} \label{eq:pseudo-interference}
\widetilde{\ICI}_l = \Re \left\{  \text{\scriptsize $ \frac{ \ICI[2l]-\ICI[2l+1]}{2}$}\right\} +\jrm
 \Im \left\{  \text{\scriptsize $  \frac{\ICI[2l+1]-\ICI[2l+2]}{2}$}\right\}
\end{equation} 
and
\begin{equation} \label{eq:CombinedNoise} 
\widetilde{\rm n}_l = \Re \left\{  \text{\scriptsize $  \frac{ n[2l]-n[2l+1]}{2}$}\right\} +\jrm
 \Im \left\{  \text{\scriptsize $  \frac{n[2l+1]-n[2l+2]}{2}$}\right\}. 
\end{equation}

The interpretation of~\eqref{eq:rtilde1[l]} is exact only for a ZF equalizer, for which $\widetilde{\rm ICI}_n=0$. Otherwise, the post-combined ICI term $\widetilde{\rm ICI}_l$ is not purely interference, but in fact also contains a desired-signal component. 
Taking  this into account, 
and as shown in Appendix B-C., 
the combined received symbols can be expressed more precisely as follows. 
\begin{Lem} \label{lem:RO-QPSK_Rxsig} 
The post-combined received symbols in \eqref{eq:rtilde[l]} or \eqref{eq:rtilde1[l]}  can be expressed in term of the QPSK symbol \eqref{eq:RO-QPSKqpsksymb}  as 
\begin{equation} \label{eq:rtilde2[l]}
\tilde{r}[l] 
= \mu_{w,G}  q_l +\mathcal{I}_l+\widetilde{\rm n}_l
\end{equation}
where 
\begin{equation} 
\mu_{w,G}  = \frac{1}{\Nsc}\sum_{k=0}^{\Nsc-1}w_k G_k  \label{eq:nuEH} 
\end{equation} and $\Ical_l$ is an interference term uncorrelated with $q_l$.
\end{Lem}
To be more precise, the ICI term $\Ical_l$ is uncorrelated but not independent of $q_l$; however, its real part is independent of $\alpha_{2l}$ and its imaginary part is independent of $\alpha_{2l+1}$.

From Lem.~\ref{lem:RO-QPSK_Rxsig}, we obtain the following exact SINR expression, proved in Appendix B-D. 
\begin{Prop} \label{lem:SINR_RO-QPSK}
The effective SINR of RO-QPSK transmission~\eqref{eq:rtilde2[l]} with i.i.d. I/Q components $\{ \alpha_m\}_{m=0}^{\Nsc-1}$  is 
\begin{equation}
\SINR^{\rm RO-QPSK} = \frac{\mu^2_{w,G}}{
\varrho^2_{w,G} + \frac{1}{2} \mu_{w,|E|^2}},
\end{equation} 
where $\mu_{w,G}$ is given in~\eqref{eq:nuEH}, the noise power is 
\begin{equation}
\mu_{w,|E|^2}= 
\frac{1}{\Nsc} \sum_{k=0}^{\Nsc-1} w_k |E_k|^2
\end{equation}
and the interference power is  
\begin{equation}
\varrho^2_{w,G} = \nu_{w,G}-\mu_{w,G}^2  
\end{equation}
with 
\begin{eqnarray}
\nu_{w,G}  
&=&  \frac{1}{2\Nsc} 
\sum_{k=0}^{\Nsc-1} w_k G_k\left( w_k G_k + (2-w_k)G_{(\frac{\Nsc}{2}-k)} \right) \nonumber 
 \\
 &\approx & \frac{1}{\Nsc} \sum_{k=0}^{\Nsc-1}w_k G^2_k \label{eq:nuwGapprox}
\end{eqnarray}
and therefore $\varrho^2_{w,G}\approx \sigma^2_{w,G}$.
\end{Prop}

Compared to QPSK, the quantities in this SINR expression are weighted means over the subcarrier coefficients, with Hann weights $w_k$.
The signal power is the square of the weighted mean of the equalized subcarrier channels, and the noise power is the weighted average power of the equalizer coefficients. 
Since  $ \nu_{w,G} \approx  \mu_{w,G^2}$ in~\eqref{eq:nuwGapprox}, 
the interference power is approximately the weighted variance of the equalized per-subcarrier channels.

\begin{figure}[t]
\vspace{-0.0cm} 
 \centering
\includegraphics[width=.49\textwidth]{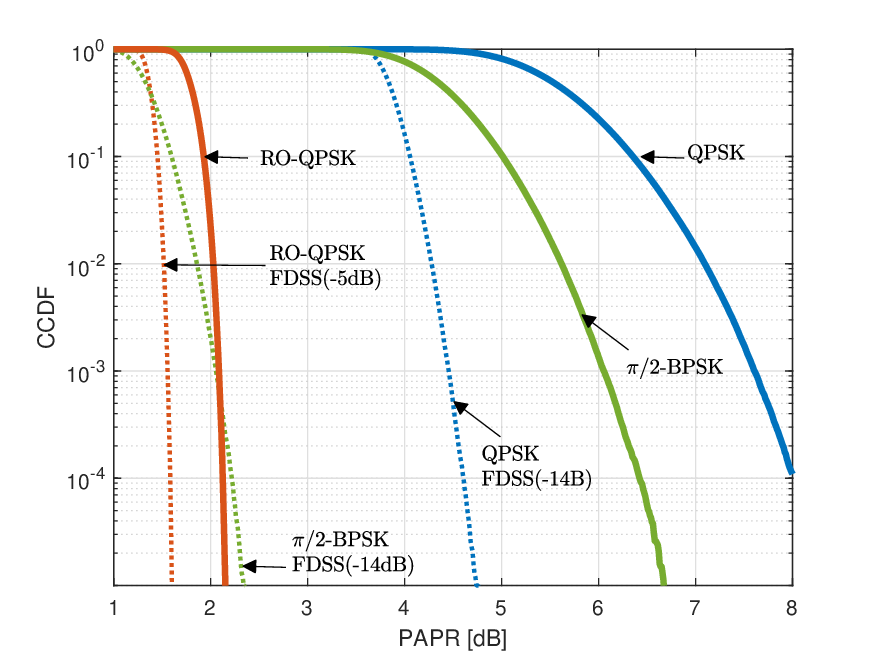}
\vspace{-0.5cm}
	\caption{PAPR comparison. \label{fig:PAPR}}
\vspace{-0.3cm}
\end{figure}

\section{Performance Analysis}
In this section, we provide a performance analysis of the transmitted signal characteristics, together with a theoretical end-to-end performance verification and evaluation of the derived effective SINRs.

\subsection{Transmit Signal Characteristics}
\subsubsection{PAPR}
Given that the average transmit power is set to $\expect{|s[n]|^2} = \frac{\Nsc}{\Nfft} $, the PAPR of the OFDM symbol $s[n]$ is 
\begin{equation}  
{\rm PAPR } =  \frac{\Nfft  }{\Nsc}\max_{0\leq n<\Nfft} |s[n]|^2 .
\label{eq:PAPR_Exp}
\end{equation} 

Fig.~\ref{fig:PAPR} shows the complementary cumulative distribution function (CCDF) of the PAPR for the considered constellations with $\Nsc =96$ and $\Nfft = 2048$.  Even with DFT-s-OFDM, the maximum PAPR of $\pi/2$-QPSK and QPSK remains relatively high: about 6 and 7.5 dB at $10^{-3}$ CCDF, respectively. 
This can be significantly reduced with FDSS. FDSS(-14dB) with $\pi/2$-BPSK results in a very low maximum PAPR, slightly above  2dB. Applying the same window to QPSK can bring the PAPR down to 4.5 dB at $10^{-3}$ CCDF.  

RO-QPSK provides directly a maximum PAPR of about 2 dB, comparable to $\pi/2$-BPSK with FDSS(-14dB) at $10^{-3}$ CCDF. The PAPR of RO-QPSK can be further reduced using FDSS, for instance, considering FDSS(-5dB)  results in maximum PAPR of 1.7 dB. 

Since FDSS(-14dB) with $\pi/2$-BPSK  provide comparable maximum PAPR to RO-QPSK, we will focus on this window for  $\pi/2$-BPSK. 
The PAPR and error-rate performances of intermediate FDSS attenuation would lie in between the no-FDSS and FDSS(-14dB) cases.

\begin{figure}[t]
\centering
\vspace{-0.0cm} 
\includegraphics[width=.5\textwidth]{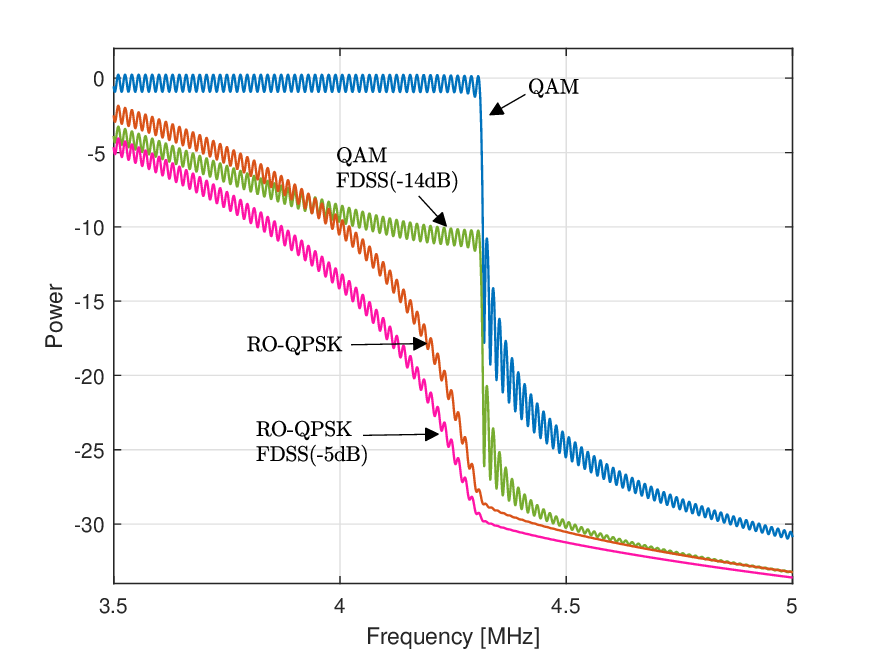}
\vspace{-0.5cm}
	\caption{Average OOB emission. \label{fig:OOB}} 
\vspace{-0.3cm} 
\end{figure}

\subsubsection{OOB Emission}

Spectrum shaping can reduce out-of-band (OOB) emission.  While this reduction alone is generally insufficient to meet regulatory requirements, it can relax the design constraints of additional filtering for OOB suppression. The strongly shaped spectrum of RO-QPSK is also beneficial from this perspective.

Given a CP-OFDM symbol duration of $T=T_{\rm S}+T_{\rm CP}$ where
$T_{\rm S}= 1/f_{\Delta {\rm sc}}$  is the useful OFDM symbol length corresponding to the  subcarrier spacing $f_{\Delta {\rm sc}}$, and $T_{\rm CP}$ is the CP duration, 
the average power spectrum of OFDM, $P(f)$  at frequency $f$, can be computed in general as in~\cite{VanDeBeekComLet09} based on the covariance of the subcarrier coefficients.  Assuming independent  subcarrier coefficients with FDSS, 
\begin{equation}
P(f) =  
\eta^2 \sum_{k=0}^{\Nsc-1}  F_k^2\expect{|X_k|^2 } {\rm sinc}^2 \left(T(f-f_{\Delta {\rm sc}}k)\right)
\end{equation} 
where $\expect{|X_k|^2 } =  1 $ for i.i.d. QAM,  and $\expect{|X_k|^2 } = w_k $ for RO-QPSK following Lem.~\ref{lem:PSD_RO-QPSK}.

Fig.~\ref{fig:OOB} illustrates the resulting OOB emission  with $f_{\Delta {\rm sc}}=15$ kHz, $T_{\rm CP}= \frac{9}{128} T_{\rm S}\approx 7$\%, and $\Nsc= 288$, corresponding to a signal bandwidth of 4.32 MHz. 
As observed, conventional FDSS applied to i.i.d. QAM already yields lower asymptotic  OOB emission.
The asymptotic OOB level of RO-QPSK is similar; however,  the spectral decay toward the band edge is significantly more gradual, providing additional attenuation at the band edge. Applying FDSS on top of RO-QPSK can further accelerate the spectral decay and slightly reduce the asymptotic OOB emission.

\begin{figure}[t]
\centering
\vspace{-0.0cm} 
\includegraphics[width=.5\textwidth]{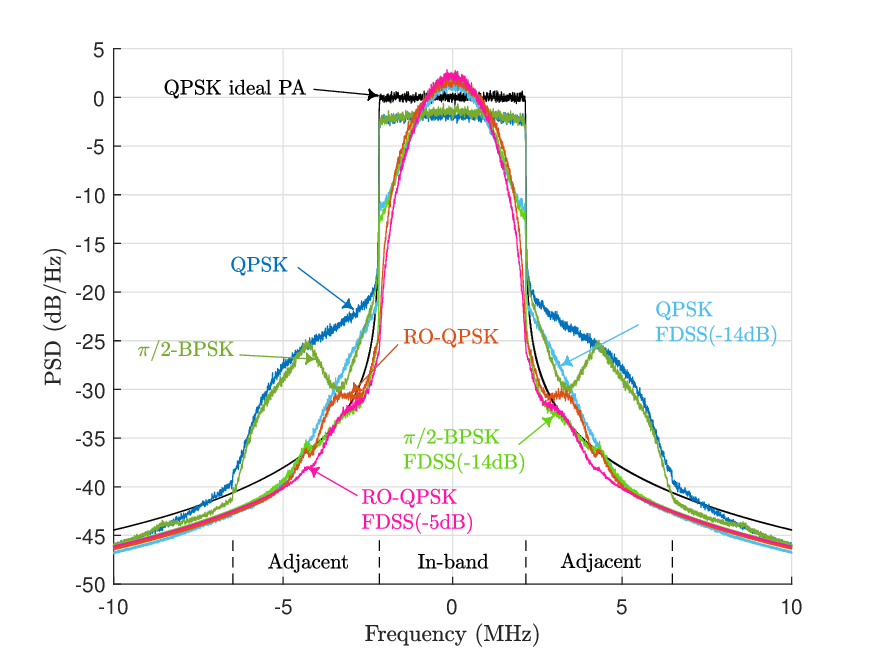}
\vspace{-0.5cm}
	\caption{PSD with nonlinear PA at saturation level (${\rm IBO}= 0$ dB). \label{fig:PSD}} 
\vspace{-0.3cm} 
\end{figure}

\begin{figure}[t]
\vspace{-0.0cm} 
\subfigure[ACLR \label{fig:ACLR}]{\includegraphics[width=.5\textwidth]{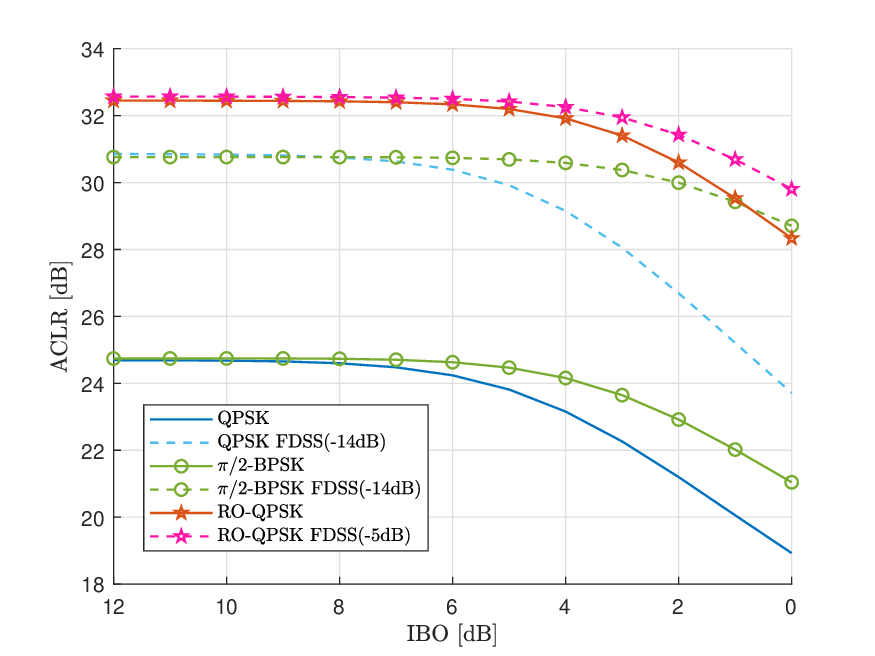}}
\subfigure[EVM\label{fig:EVM}]{\includegraphics[width=.5\textwidth]{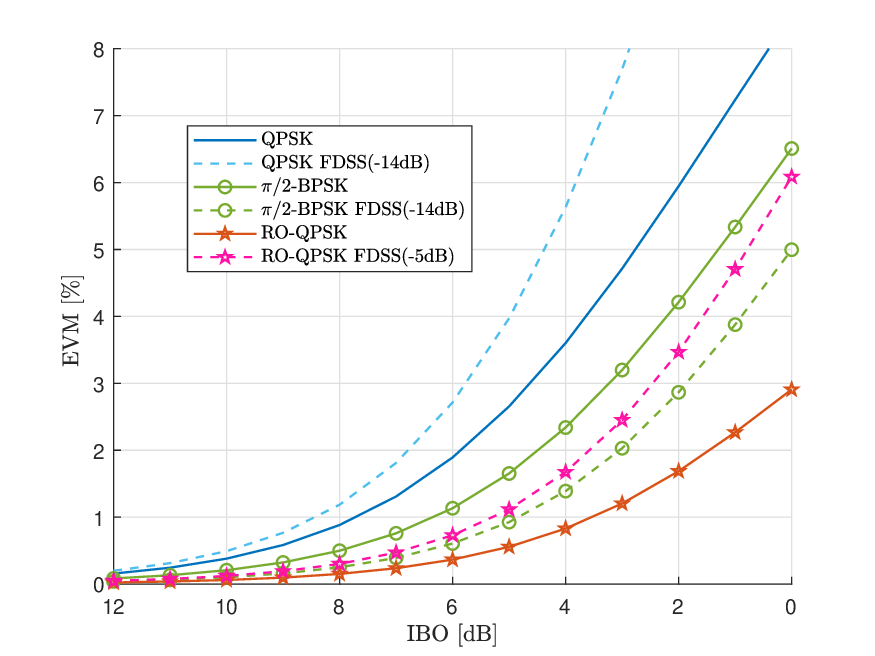}}
\vspace{-0.2cm}
	\caption{ACL and EVM with nonlinear PA. \label{fig:ACLREVM}}
\vspace{-0.3cm}
\end{figure}

\subsubsection{Spectral Regrowth with Nonlinear Power Amplifier (PA)} 
Reduced transmit signal envelope fluctuations, as reflected by a lower PAPR, typically correspond to improved operation in the presence of nonlinear PA distortion~\cite{hossain2019nonlinear,hossain2020dft}. 
The resulting PA efficiency gains depend on the PA characteristics and bandwidth allocation.   We show an example here for the widely-used  memoryless  Rapp model~\cite{Rapp1991HPA}  with $p=2$, commonly adopted for nonlinear solid-state power amplifier characterization.   
The signal amplitude at the output of the PA is distorted as  $A(r) = \frac{r}{\left(1+ \left( \frac{r}{A_{\rm sat}} \right)^{2p}\right)^{\frac{1}{2p}}}$ where $r$ is input amplitude signal and $A_{\rm sat}$ is the PA saturation level. All signals are normalized to the same average power $P_{\rm in }$ prior to PA processing, controlled by input backoff (IBO) defined as $10 \log_{10} \left( A_{\rm sat}^2/P_{\rm in } \right) $ [dB].

In Fig.~\ref{fig:PSD}, we show the power spectral density (PSD) after PA with ${\rm IBO} =0$ [dB], i.e. when the signal power is at saturation level, thus affected by strong non-linearity resulting in 
large spectral regrowth. Again, we consider a signal with $f_{\Delta {\rm sc}}=15$ kHz and $\Nsc= 288$, corresponding to a signal bandwidth of 4.32 MHz. 
The y-axis is scaled  
such that  0 dB  corresponds to the inband level of the reference QPSK scheme with ideal PA. 
The PSD is obtained using Welch's averaged periodogram method over a continuous sequence of 500  independently generated CP-OFDM symbols. Each OFDM symbol is  made of $\Nfft = 2048$ samples with a CP length of $\Ncp = 288$. Periodograms are computed using a Hamming window of length $2(\Nfft+\Ncp)$ with an FFT of 4 times the window length, and $75\%$ overlap between consecutive segments. 

As it can be seen compared to QPSK with ideal PA, nonlinear PA operation results in spectral regrowth into adjacent bands and as a by-product reduced power level in the in-band.  This is slightly improved with $\pi/2$-BPSK. The in-band power reduction for plain  QPSK and  $\pi/2$-BPSK is of about 2dB which will directly translate in reduced link SNR. 
Applying FDSS significantly reduced the spectral regrowth, notably with $\pi/2$-BPSK. With RO-QPSK, We observe a PSD similar to that of $\pi/2$-BPSK with FDSS(-14dB) but with slightly larger shoulders.  
Using FDSS(-5dB) with RO-QPSK further suppresses the spectral shoulders and yields the best spectral containment among the schemes.

To quantify the spectral containment, we compute the adjacent channel leakage ratio (ACLR) as the ratio of the integrated  power within the in-band over the integrated leaked power within adjacent bands of same size (as indicated on Fig.~\ref{fig:PSD}).   
The results are shown on Fig.~\ref{fig:ACLR} as function of IBO. Overall, the ACLR performance is consistent with PAPR comparison. RO-QPSK provides ACLR comparable to, and generally higher than, $\pi/2$-BPSK with FDSS(-14dB) over a wide range of IBO values. 
However, when getting closer to saturation level ${\rm IBO} <1$ dB, $\pi/2$-BPSK with FDSS(-14dB) starts to be slightly better, consistent with  the higher RO-QPSK `PSD's shoulders' observed in Fig.~\ref{fig:PSD}.

In Fig.~\ref{fig:EVM}, the different schemes are also compared in terms of error vector magnitude (EVM) between the demodulated constellation symbols  and the transmitted constellation symbols after non-linear PA. 
FDSS compensation and DFT de-spreading is applied in the demodulation. Prior to EVM computation, deterministic average gain compression and phase offset introduced by the PA are removed using least-squares estimation in order to measure only residual nonlinear distortion. 
While FDSS reduces the EVM of $\pi/2$-BPSK, the same trend is not observed for QPSK and RO-QPSK. 
This indicates that the reduction in envelope fluctuation enabled by FDSS does not necessarily translate into a lower post-demodulation EVM, suggesting a trade-off between transmitter-side and receiver-side effects. 
RO-QPSK consistently achieves the lowest EVM among all considered schemes across the consider IBO range.

Overall, RO-QPSK reduces spectral regrowth with improved ACLR and EVM, enabling operation closer to PA saturation with reduced input back-off. 

\subsection{Theoretical End-to-End  Performance Verification}

\subsubsection{Channel Models} We use 3GPP channel models~\cite{3GPP_TR_38811,3GPP_TR_38901}. 
We focus on the baseline LOS channel model NTN-TDL-C with a delay scaling of 3.5 ns. This corresponds to the mean delay spread of LOS rural scenario at S-band with a 30° elevation angle, which corresponds to the edge of a low earth orbit (LEO) satellite beam~\cite{3GPP_TR_38811}. This channel exhibits very limited frequency selectivity.
The  more frequency-selective channel NLOS  NTN-TDL-A channel is also considered with a delay scaling of 100 ns. This represents a large delay spread for satellite modeling, corresponding to the mean delay spread at a 30° elevation angle  in NLOS dense-urban scenario in S-band,  according Table 6.7.2-2a in~\cite{3GPP_TR_38811}. 
For comparison with terrestrial networks, we also consider the standard TDL-C channel with 300 ns delay scaling,  which corresponds to the mean delay spread in urban scenarios~\cite{3GPP_TR_38901}.

The LOS path components are assumed Rician-faded, while the NLOS components are Rayleigh-faded, both with Jakes' Doppler spectrum. Unless otherwise specified, the maximum Doppler shift is set to 11 Hz for TN channels and 200 Hz for NTN channels. These values are consistent with typical 3GPP assumptions for pedestrian users. For NTN, the latter corresponds to a target residual frequency offset accuracy of $\pm 0.1$ ppm after initial frequency synchronization~\cite{3gpp.ts.38.101-5.17.1.0,ErkaiPIMRC24}, assuming a $2$GHz carrier frequency, and without accounting for further refinement by receiver-side tracking loops. 
 
\subsubsection{Uncoded BER}

\begin{figure}[t]
\centering
\vspace{-0.0cm} 
\includegraphics[width=.5\textwidth]{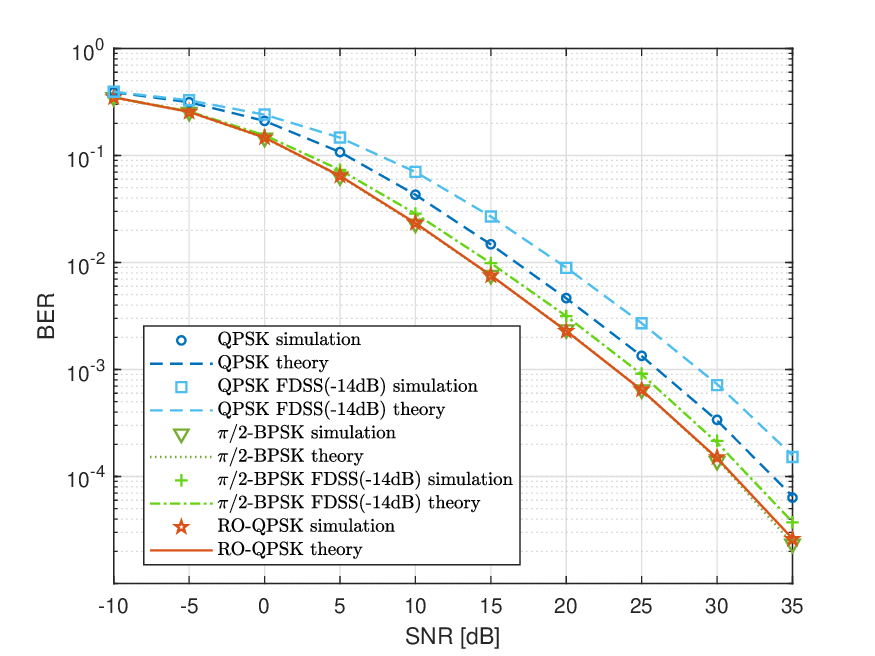}
\vspace{-0.5cm}
	\caption{Uncoded BER in NTN-TDL-C channel. \label{fig:unBER_8PRB_MMSE_NTNTDLC}}
\vspace{-0.3cm}
\end{figure}

\begin{figure}[t]
\vspace{-0.0cm} 
\includegraphics[width=.5\textwidth]{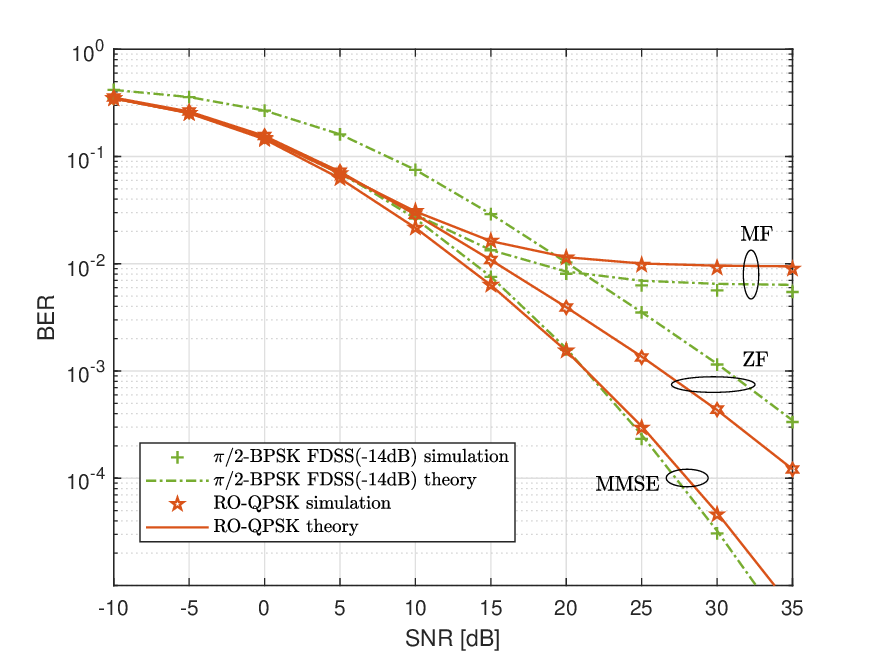}
\vspace{-0.5cm}
	\caption{Uncoded BERs for different equalizers in NTN-TDL-A channel. \label{fig:unBER_8PRB_MMSE_NTNTDLA_differentEqualizer}}
\vspace{-0.3cm}
\end{figure}

\begin{figure*}[t]
\vspace{-0.0cm} 
\subfigure[NTN-TDL-C \label{fig:Capacity_8PRB_NTNTDLC30}]{\includegraphics[width=.5\textwidth]{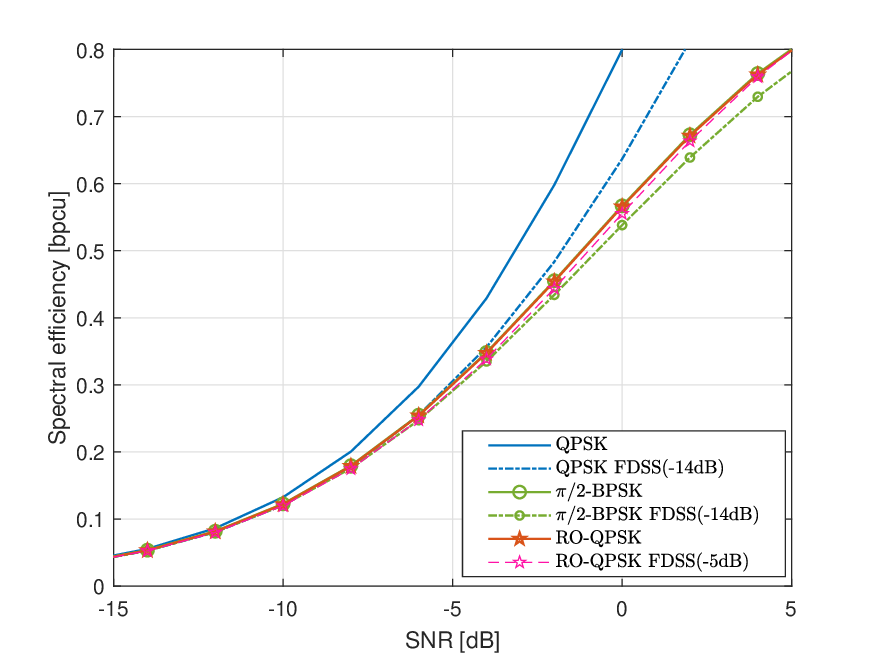}}
\subfigure[TDL-C \label{fig:Capacity_24PRB_TDLC300}]{\includegraphics[width=.5\textwidth]{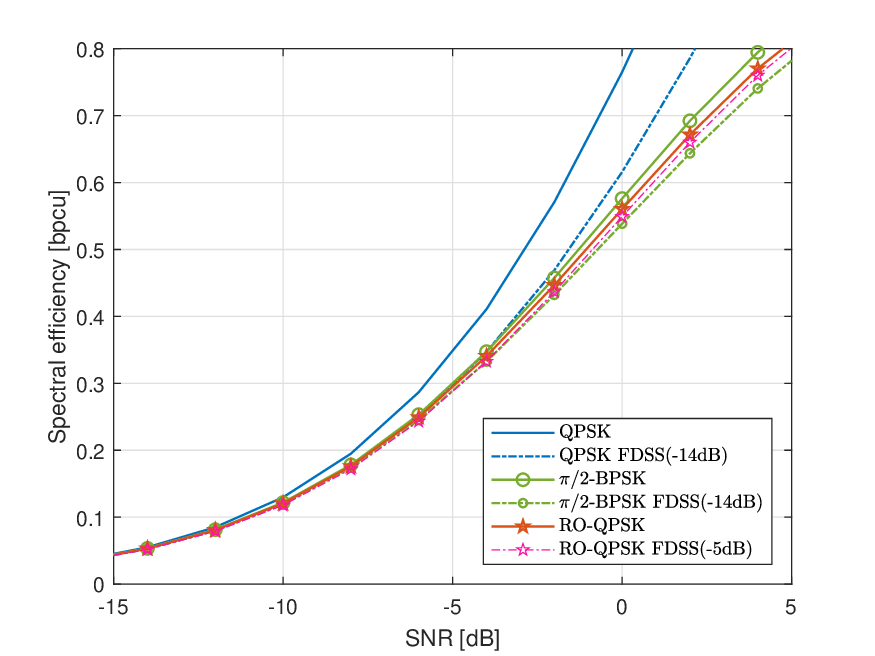}}
\vspace{-0.2cm}
	\caption{Capacity of DFT-s-OFDM with different constellations. \label{fig:Capacity}}
\vspace{-0.3cm}
\end{figure*}

To verify the derived SINR expressions in  
Eq.~\eqref{eq:r[m]} of Prop.~\ref{lem:SINR}, Eq.~\eqref{eq:rR[m]} of  Prop.~\ref{lem:SINR_BPSK}, and  Eq.~\eqref{eq:rtilde2[l]} of Prop.~\ref{lem:SINR_RO-QPSK},
Figs.~\ref{fig:unBER_8PRB_MMSE_NTNTDLC} and~\ref{fig:unBER_8PRB_MMSE_NTNTDLA_differentEqualizer} compare the simulated uncoded bit error rate (BER) for $\Nsc =96$ with a  corresponding semi-analytical BER. Assuming the interference is Gaussian, the theoretical BERs are given by $\expect{Q\left(\sqrt{ \SINR^{\rm{iid}\mathbb{C}}} \right)}$, $\expect{Q\left(\sqrt{ 2\SINR^{\rm BPSK}} \right)}$, and $\expect{Q\left(\sqrt{ \SINR^{\rm RO-QPSK}} \right)}$, for QPSK, $\pi/2$-BPSK and RO-QPSK, respectively, where the Q-function is the tail distribution function of the standard normal distribution, i.e. $Q(x)= \frac{1}{\sqrt{2\pi}} \int_x^\infty e^{-\frac{u^2}{2}} du$. The BER is averaged over block-faded channel realizations using Monte Carlo simulations. 

All curves in Fig.~\ref{fig:unBER_8PRB_MMSE_NTNTDLC} correspond to the NTN-TDL-C channel with MMSE equalization. A perfect match is observed between theory and simulation. RO-QPSK and $\pi/2$-BPSK (without FDSS) exhibit identical BER performance, while QPSK suffers an approximately 3 dB SNR loss at $10^{-3}$ BER. With FDSS(-14dB), $\pi/2$-BPSK shows a 1.5 dB loss, whereas QPSK reaches a loss of 5.5 dB.

Fig.~\ref{fig:unBER_8PRB_MMSE_NTNTDLA_differentEqualizer} considers the NTN-TDL-A channel with MMSE, ZF and MF equalization. Theoretical and simulated BERs match perfectly for MMSE and ZF, while a small discrepancy is observed for MF. We numerically verified the derived SINR expressions, and we attribute this mismatch to the ICI not being perfectly Gaussian.
Due to the higher frequency selectivity of NTN-TDL-A, the BER of RO-QPSK with MMSE is only slightly better than that of $\pi/2$-BPSK with FDSS(-14dB) at low SNR, and slightly worse above 20 dB SNR. With ZF, RO-QPSK performs significantly better than $\pi/2$-BPSK with FDSS(-14dB), whereas the opposite holds for MF.    
Overall, MMSE equalization provides the  best performance than  ZF, followed by ZF and MF. Hence, MMSE equalization is used in the following evaluations.  

\subsubsection{Capacity} 
Fig.~\ref{fig:Capacity}  evaluates the capacity, i.e., the maximum achievable spectral efficiency for the considered constellations, using the derived channel gain, noise, and interference powers. This is computed from the classical  mutual information of BPSK input   assuming Gaussian interference, and averaged over random block-faded channel realizations (see Appendix~\ref{app:MutInf} for details).

Fig.~\ref{fig:Capacity_8PRB_NTNTDLC30} shows the capacity for the NTN-TDL-C channel with $\Nsc= 96 $. All constellations perform similarly as $\snr \to 0$, while differences are increasing at higher SNR, corresponding to higher spectral efficiencies. QPSK starts to clearly outperform others from about 0.1 bpcu, while FDSS-QPSK remains aligned with first-order modulations over a wider range, though at the cost of modest PAPR reduction.  
$\pi/2$-BPSK and RO-QPSK achieve similar spectral efficiency, but RO-QPSK offers significantly lower PAPR. When FDSS(-14 dB) is applied to $\pi/2$-BPSK for comparable PAPR, its spectral efficiency remains about 1 dB lower than that of RO-QPSK. Furthermore, RO-QPSK with moderate FDSS(-5 dB) achieves still a higher spectral efficiency while achieving an even lower PAPR.

Fig.~\ref{fig:Capacity_24PRB_TDLC300} considers a wider band ($\Nsc= 288$) with the TN  TDL-C channel that exhibits higher frequency selectivity. In this case, RO-QPSK shows a clear performance loss compared to $\pi/2$-BPSK (without FDSS) of 0.6 dB. Nevertheless, $\pi/2$-BPSK with  FDSS(-14 dB) experiences an even larger maximum SNR loss of 1.4 dB.

\section{Block Error Rate Evaluations}

  \begin{figure*}[t]
\vspace{-0.0cm} 
\subfigure[$\Ninfo= 100$ bits \label{fig:BLER_2PRB_100bits}]{\includegraphics[width=.5\textwidth]{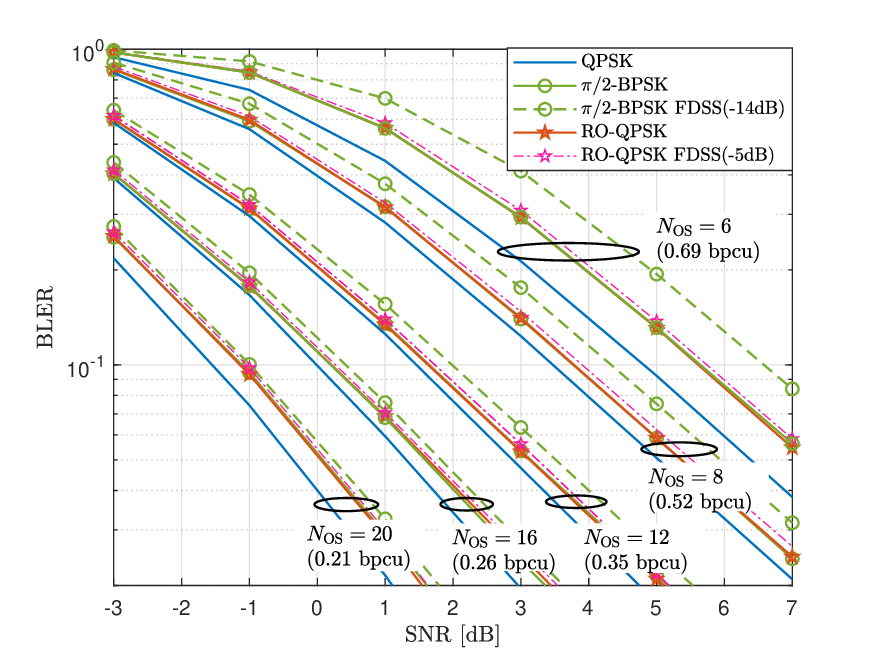}}
\subfigure[$\Ninfo = 1000$ bits  \label{fig:BLER_2PRB_1000bits}]{\includegraphics[width=.5\textwidth]{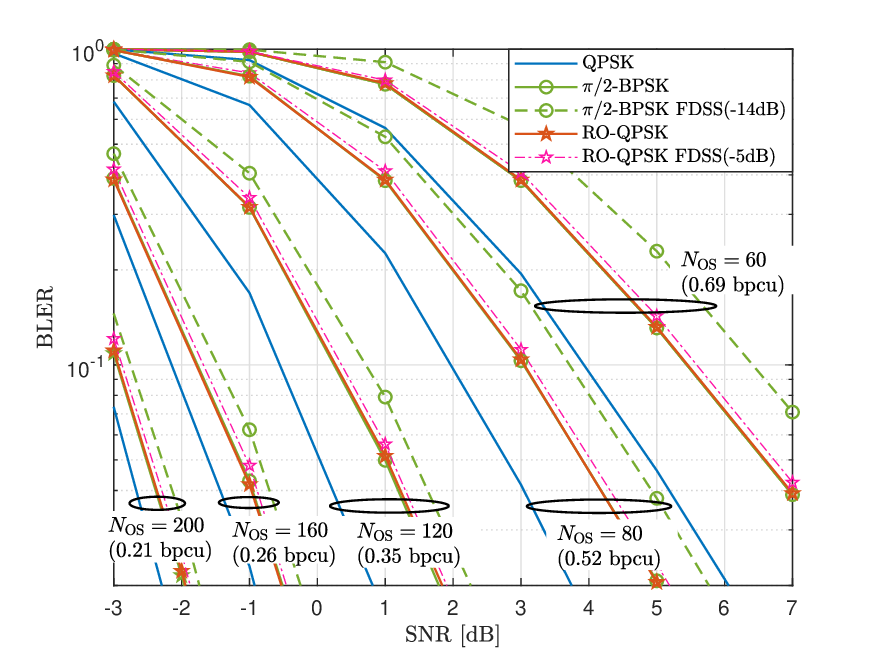}}
\vspace{-0.2cm}
	\caption{BLER in narrowband ($\Nsc = 24 $)  and NTN-TDL-C channel \label{fig:BLER_2PRB_NTNTDLC}}
\vspace{-0.3cm}
\end{figure*}

\subsection{Assumptions}
\subsubsection{Resource Allocation} We consider a transmission with $\Ninfo$ information bits, encoded and modulated over $\Nsc$ subcarriers   and $\Nos$ OFDM symbols. The spectral efficiency is thus given by $\frac{\Ninfo}{\Nsc\Nos}$ bpcu. The OFDM subcarrier spacing is 15 kHz,  $\Nfft = 1024$ and $\Ncp=72$ is used. 

\subsubsection{Genie-aided Channel Estimation} 
We use an almost perfect channel estimation, except for the fact that it is ICI-corrupted due to Doppler spread.  
The channel estimate is obtained by transmitting 
an OFDM-modulated pilot sequence of length $\Nsc$. On subcarrier $k$, a pilot symbol $P_k$ is received according to~\eqref{eq:Yk} as 
\begin{equation} \label{eq:YkP}
Y^{(P)}_{k} = \eta\sqrt{\snr} F_{k} H_{k,k}P_k +I_k^{(P)}
\end{equation}
 where $I_k^{(P)}= \sum_{k'= 0, k' \neq k}^{\Nsc-1}\eta \sqrt{\snr}  F_{k'}  H_{k,k'} P_{k'}$ represents ICI. The channel estimation  on subcarrier $k$ is thus
\begin{equation} 
\widehat{\tilde{H}_k} =  Y^{(P)}_{k}/P_k = \eta\sqrt{\snr} F_{k} H_{k,k} + I_k^{(P)}/P_k .
\end{equation}
This gives a perfect channel estimate   $\widehat{\tilde{H}_k} = \tilde{H}_k$ only if the fading is purely constant over the OFDM symbol duration.  For evaluations, we use a Zadoff-Chu sequence of length $\Nsc$. Such sequences exhibit a higher PAPR than the target levels considered in this work. Low-PAPR pilot sequences based on DFT-precoded $\pi/2$-BPSK (computer-generated or  Gold sequences)  were introduced in 5G Rel. 16~\cite{3gpp.38.211.16.6.0}; however, they exhibit magnitude fluctuations across  subcarriers that may degrade channel estimation quality. The design of low-PAPR pilot sequences for channel estimation is beyond the scope of this paper and is left for future work.

\subsubsection{Forward Error Correction (FEC) Coding and Decoding}
Information bits are encoded using low density parity check (LDPC) coding prior to modulation. Comparisons are performed under the same spectral efficiency, such that the code rate is halved for QPSK. 

Based on the channel estimate $\widehat{\tilde{H}_k}$, the corresponding effective channel gain and SINR are computed, and the input LLR values at the LDPC decoder are scaled accordingly. Specifically, according  to Eq.~\eqref{eq:r[m]} of Prop.~\ref{lem:SINR}, Eq.~\eqref{eq:rR[m]} of  Prop.~\ref{lem:SINR_BPSK}, and  Eq.~\eqref{eq:rtilde2[l]} of Prop.~\ref{lem:SINR_RO-QPSK}, we have
\begin{align} 
&{\rm LLR}([b_{2m}, b_{2m+1}] | r[m]) \!=\!  \textstyle{ \frac{ \SINR^{\rm{iid}\mathbb{C}} }{\mu_{G}} } 2\sqrt{2} [\Re \{ r[m]\},\, \Im\{ r[m]\}] \nonumber\\
&{\rm LLR}(b_m | r_{\rm R}[m]) =   \textstyle{ \frac{  \SINR^{\rm BPSK}}{\mu_{G}}}  2\, r_{\rm R}[m]\\
&{\rm LLR}([b_{2l}, b_{2l+1}] | \tilde{r}[l]) =    \textstyle{\frac{  \SINR^{\rm RO-QPSK} }{ \mu_{w,G}}}   2\sqrt{2}[\Re \{ \tilde{r}[l]\},\, \Im\{ \tilde{r}[l]\}]\nonumber
\end{align}
for QPSK, $\pi/2$-BPSK, and RO-QPSK, respectively. Finally, LDPC decoding with 4 iterations is applied.

\subsection{BLER Results}
\subsubsection{Narrowband LOS NTN}

Fig.~\ref{fig:BLER_2PRB_NTNTDLC} shows the BLER in the NTN-TDL-C channel with 100- and 1000-bit payloads in a narrowband transmission with $\Nsc =24$, and various spectral efficiencies obtained by selecting different $\Nos$.  
In low spectral efficiencies, all modulations provide similar performance, and the differences increase with higher spectral efficiencies. With a longer information block, the performance shifts to lower SNRs, and performance differences also become more pronounced. 
While QPSK consistently outperforms the others constellations in BLER, with up to 1.4 dB SNR gain at $10^{-1}$ BLER over $\pi/2$-BPSK, the  lower PAPR of order-one constellations ($\pi/2$-BPSK and RO-QPSK) can, in practice, yield overall better link performance than QPSK for low or moderate spectral efficiencies. RO-QPSK and $\pi/2$-BPSK (without FDSS) achieve similar BLER performance, although RO-QPSK provides a much lower PAPR.  $\pi/2$-BPSK with FDSS(-14dB), which offers similar PAPR, incurs an additional SNR loss of up to 1 dB at $10^{-1}$ BLER. Applying  FDSS(-5dB) to  RO-QPSK for further PAPR reduction causes only  a minor SNR loss.

\subsubsection{Frequency-Selective Channels}
\begin{figure}[t]
\vspace{-0.0cm} 
\centering
\includegraphics[width=.5\textwidth]{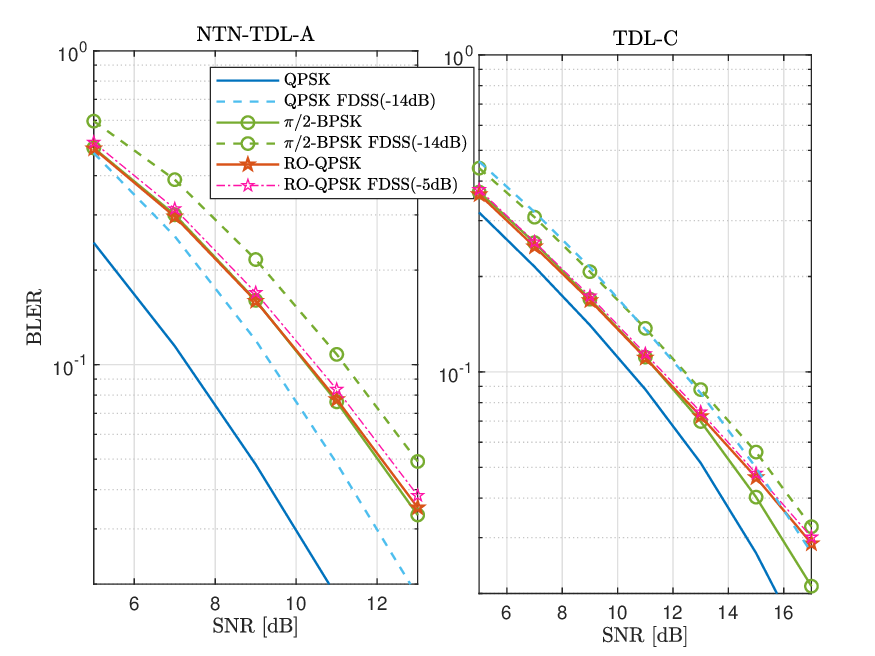}
\vspace{-0.5cm}
	\caption{BLER  at $0.69$  [bpcu] in narrowband  ($\Nsc = 24 $) and more frequency-selective channels. Left: NTN-TDL-A, right: TDL-C. \label{fig:BLER_freqSelective}}
\vspace{-0.3cm}
\end{figure}

Fig.~\ref{fig:BLER_freqSelective} compares the BLER performance  for the same narrowband case ($\Nsc =24$) under more frequency-selective channels: NTN-TDL-A(100ns) and TN TDL-C(300ns). By numerical evaluations, the $0.9$-correlation coherence bandwidth of these two channel models is found to be 48 and 22 subcarrriers, respectively; and thus this narrowband allocation is already slightly larger than the coherence bandwidth of the TDL-C channel. 
For comparison, we focus on the high spectral efficiency of $0.69$ bpcu with a 1000-bit payload, where the performance gaps are the largest. The figure also includes the performance of QPSK with FDSS(-14dB). $\pi/2$-BPSK with FDSS(-14dB) shows a smaller SNR loss than in the NTN-TDL-C channel. Notably, with TDL-C, the performances of all  schemes are closer, and RO-QPSK starts to deviate from $\pi/2$-BPSK (without FDSS) in low-BLER regime, but still outperforms  $\pi/2$-BPSK with FDSS(-14dB). 

\begin{figure}[t]
\vspace{-0.0cm} 
\centering
\includegraphics[width=.48\textwidth]{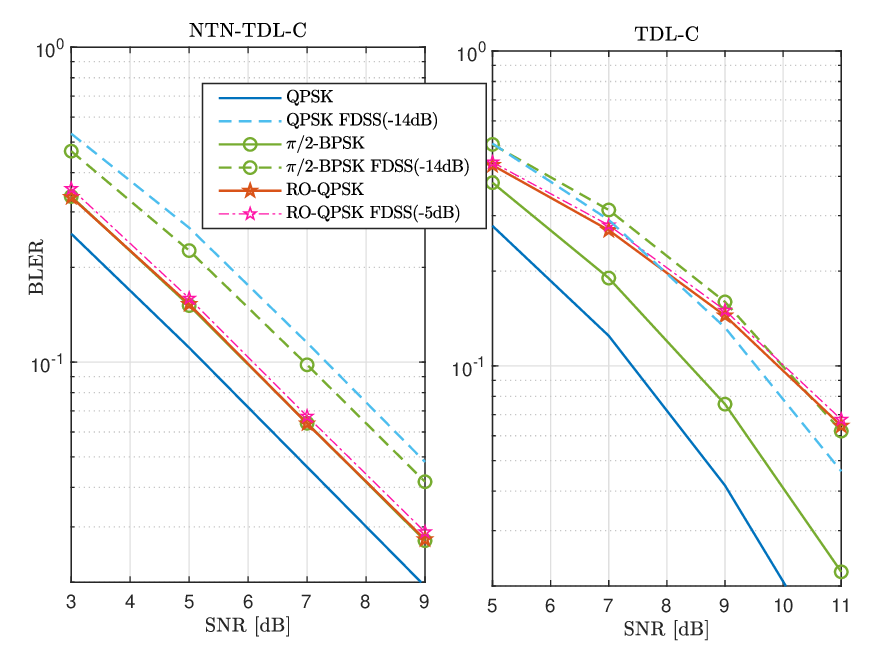}
\vspace{-0.2cm}
	\caption{BLER at $0.69$  [bpcu] in a wider band ($\Nsc = 288$). Left: NTN-TDL-C, right: TDL-C. \label{fig:BLERwiderband}}
\vspace{-0.3cm}
\end{figure}

Fig.~\ref{fig:BLERwiderband} compares BLER in a wider band ($\Nsc =288$) to better show the impact of frequency selectivity, considering NTN-TDL-C  with low selectivity and TDL-C with high selectivity. This band allocation is ten times larger than the coherence bandwidth of the TDL-C channel.   Again, $0.69$ bpcu with a 1000-bit payload ($\Nos =60$) is considered. 
With NTN-TDL-C, the wideband performance is similar to the narrowband case and again RO-QPSK provides a clear improvement over $\pi/2$-BPSK with FDSS, still tight with the performance of $\pi/2$-BPSK without FDSS. However, with TDL-C,  the performance of both RO-QPSK and $\pi/2$-BPSK with FDSS(-14dB), relative to $\pi/2$-BPSK without FDSS, degrades significantly, and RO-QPSK performs similarly than to $\pi/2$-BPSK with FDSS(-14dB).

\begin{figure}[t]
\vspace{-0.0cm} 
\centering
\includegraphics[width=.5\textwidth]{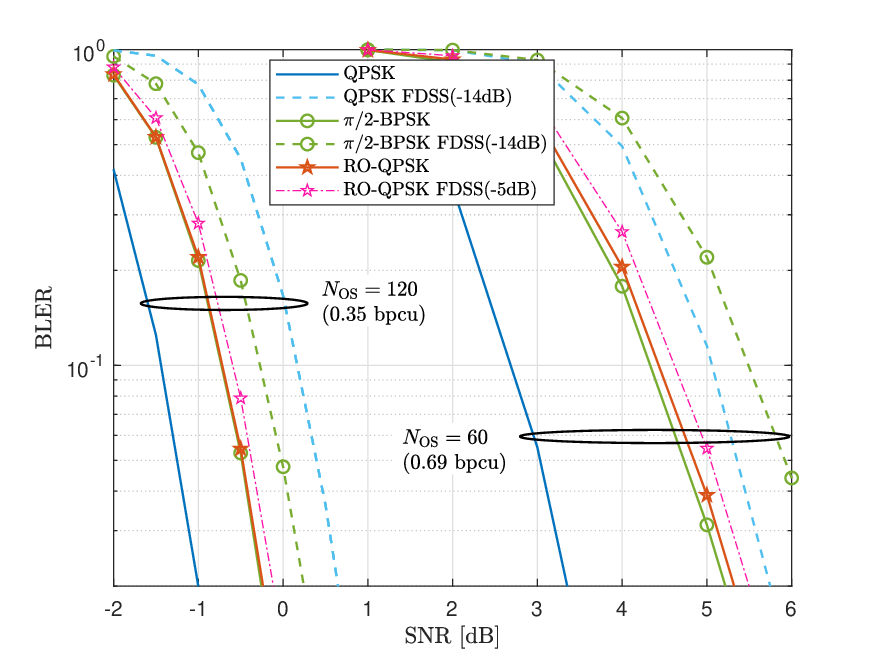}
\vspace{-0.6cm}
	\caption{BLER with large  residual Doppler spread. Narrowband ($\Nsc = 24 $) in NTN-TDL-C channel and 1000 bits payload. \label{fig:BLERdoppler}}
\vspace{-0.0cm}
\end{figure}

\begin{figure}[t]
\vspace{-0.0cm} 
\includegraphics[width=.48\textwidth]{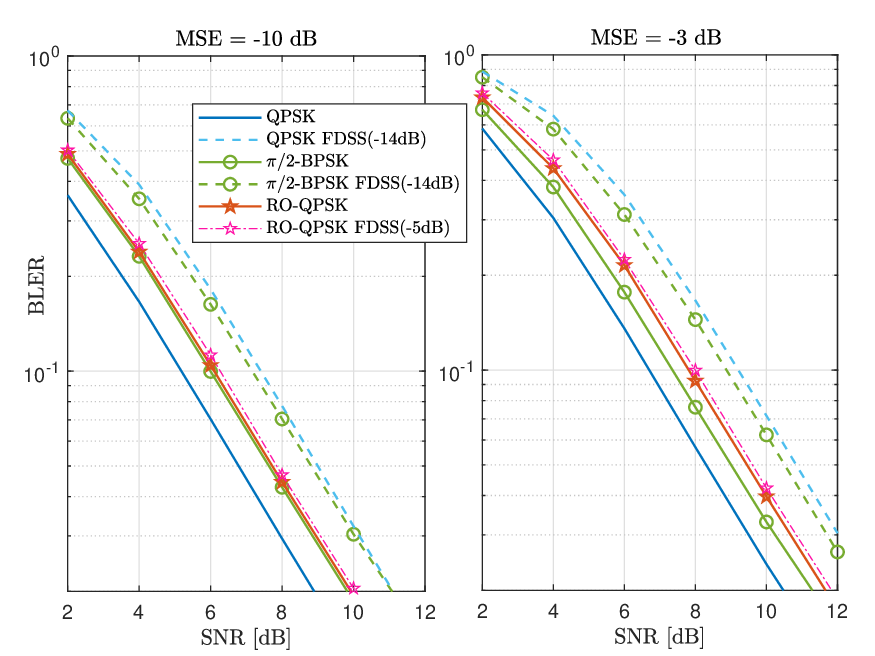}
\vspace{-0.1cm}
	\caption{BLER with noisy channel estimation. Narrowband ($\Nsc = 24 $) in NTN-TDL-C channel and 100 bits payload. \label{fig:BLERnoisy}}
\vspace{-0.3cm}
\end{figure}

\subsubsection{Large Residual Doppler and Noisy Channel Estimation}
Finally we consider impact of large residual frequency offset and imperfect channel estimation with  $\Nsc=24$ and NTN-TDL-C channel. 
  
Fig.~\ref{fig:BLERdoppler} shows the BLER  of 1000-bit payload with $\Nos = 60$ and $120$, under a maximum  Doppler of 2000 Hz, 
which is significantly larger than typical residual frequency offsets expected in practical NTN systems after synchronization.      
A large Doppler shift causes faster channel time variations, which makes accurate channel estimation more challenging, as it requires a larger pilot overhead. However, if the channel can still be estimated reliably, the performance actually improve under faster-varying channel, since then the data block  experiences a greater diversity of fading realizations, enabling the rate to approach the ergodic capacity faster.
Therefore, compared to Fig.~\ref{fig:BLER_2PRB_1000bits}, the overall performance in Fig.~\ref{fig:BLERdoppler}  improves. 
At high Doppler, non-negligible channel variations  within a single OFDM symbol also occur, creating  ICI both during the data transmission~\eqref{eq:YkP} and channel estimation~\eqref{eq:Yk}. The SINR expressions derived under the assumption of constant channel within one OFDM symbol are thus not  anymore precise. Meanwhile, for schemes with a maximum spectrum efficiency of one as considered here, i.e., operating in low or moderate SNR, the discrepancy from the true SINR in the LLR computation is tolerable, and the performances in Fig.~\ref{fig:BLERdoppler} show that  the considered schemes are in general robust under strong Doppler.  
We observe that RO-QPSK departs slightly from $\pi/2$-BPSK without FDSS having a small SNR loss of $0.1$ dB for the large spectral efficiency case of 0.69 bpcu, but still providing more than 1 dB SNR gain over  $\pi/2$-BPSK with FDSS(-14dB).

Finally, in Fig.~\ref{fig:BLERnoisy}, we evaluate  the impact of noisy channel estimation for 100-bit payload with $ \Nos = 6$, at thus 0.69 bpcu spectral efficiency, and $200$ Hz maximum Doppler. 
The channel estimation is modeled as 
\begin{equation}
\widehat{\tilde{H}_k}^{(e)}  = \widehat{\tilde{H}_k} + e_k
\end{equation}
where $e_k$ is a zero-mean Gaussian variable whose variance 
equals the channel estimation mean square error (MSE). Two cases are shown: ${\rm MSE} = -10$ and $-3$ dB. With channel estimation errors, the overall performance degrades relatively to Fig.~\ref{fig:BLER_2PRB_1000bits}. 
For QPSK, the performance degradation is 0.4 dB  and  1.9 dB SNR loss, with ${\rm MSE} = -10$ and $-3$ dB, respectively. For $\pi/2$-BPSK with FDSS(-14dB), the SNR loss is slightly larger 0.5 dB and 2.2 dB, with ${\rm MSE} = -10$ and $-3$ dB, respectively. Similarly to  $\pi/2$-BPSK with FDSS(-14dB), RO-QPSK shows a slightly larger SNR degradation. 
If the channel estimation error remains small, RO-QPSK still  performs close to $\pi/2$-BPSK without FDSS, but for larger errors, an SNR gap of 0.2 dB appears. Nonetheless, RO-QPSK maintains superior performance over $\pi/2$-BPSK with FDSS(-14dB) for a same PAPR level. Even with FDSS(-5dB) for lower PAPR, RO-QPSK continues to outperform it.

In the presence of large Doppler or imperfect channel estimation, a small performance gap may appear from plain $\pi/2$-BPSK, but the advantage over  $\pi/2$-BPSK with FDSS at equivalent PAPR remains largely preserved.

\section{Conclusion}
This paper investigated low-PAPR, low-order modulation schemes for DFT-s-OFDM transmissions operating in low-SNR regimes, with particular relevance to satellite communications and their convergence with terrestrial networks. First, a general    analytical  framework  was developed  
for DFT-s-OFDM with FDSS and any single-tap equalization. 
Closed-form SINR expressions were derived for i.i.d. complex constellations and $\pi/2$-BPSK, enabling theoretical BER and capacity evaluations.  
Then, an alternative correlation-based modulation scheme,  termed  repeated-and-offset QPSK (RO-QPSK), was proposed.  
RO-QPSK is implemented as a lightweight extension of conventional QPSK, relying solely on bit-level processing at the transmitter and simple symbol combining at the receiver. 
The pre-modulation bit encoding induces a structured correlation among transmitted symbols, resulting in an inherent spectrum shaping with a Hann-window profile and a PAPR on the order of 2 dB,  
which is significantly lower than $\pi/2$-BPSK without FDSS and  comparable to $\pi/2$-BPSK with typical 5G FDSS.
The analysis and simulations show that, in narrowband and moderately frequency-selective channels representative of many satellite scenarios, the joint transmitter-receiver design of RO-QPSK can provide a performance improvement of about 1 dB SNR compared to 5G-like approach at an equivalent PAPR target. In more frequency-selective channels, both approaches exhibit comparable performance.
 
The proposed analytical framework and modulation approach provide useful insights and tools for future work  
on improving $\pi/2$-BPSK FDSS transmissions,   
as well as for investigating symbol correlation-based or differential encoding schemes that could enable lower PAPR in  DFT-s-OFDM  
at higher  spectral efficiencies,  
 such as with higher-order QAM or amplitude and phase shift keying (APSK). 
\begin{appendices}

\section{Legacy Constellations}
\subsection{Proof of Prop.~\ref{lem:SINR} (SINR of i.i.d Complex Constellation)}
\label{app:SINR_QAM}

Starting from \eqref{eq:r[m]}, we define the SINR as 
\begin{equation}
\SINR^{\rm{iid}\mathbb{C}} = \frac{\mu_{G}^2 }{\expect{|\ICI[m]|^2}+ \expect{|n[m]|^2}}
\end{equation}
 where the expectation is taken over both the data symbols and noise samples. 

The noise power is given by direct averaging of \eqref{eq:n[m]}, leading to \eqref{eq:muE}. Assuming independent and zero-mean constellation symbols,  the interference power can be decomposed  as 
\begin{eqnarray}
\expect{ | \ICI[m]|^2} 
&=& \expect{\left| \left(\sum_{n=0}^{\Nsc-1} 
x[n]g_{m-n} \right) - g_{0}x[m]\right|^2}  \nonumber\\
&=& \mu_{G^2}-\mu_{G}^2 
\end{eqnarray}
where $\mu_{G^2} = \sum_{n=0}^{\Nsc-1}\left|g_{m-n}\right|^2 $. 
Noting hat $g_{-a} = g_{a}^*$ and by direct expansion from~\eqref{eq:gm}, 
we can generally write
\begin{equation}\label{eq:PlancherelThG}
\sum_{a=0}^{\Nsc-1}g_{a+x} g^*_{a+y} = \frac{1}{\Nsc} 
\sum_{k=0}^{\Nsc-1} G^2_k e^{\frac{2\pi}{\Nsc}k (x-y)}.
\end{equation}
Using this, we obtain
$\sum_{n=0}^{\Nsc-1}\left|g_{m-n}\right|^2= \frac{1}{\Nsc} \sum_{k=0}^{\Nsc-1}G^2_k$ 
which can be seen as a form of Plancherel theorem.

\subsection{Proof of Cor.~\ref{Cor:ZF_MMSE} (SINR of ZF of MMSE)}  
\label{app:SINR_MMSE}

For ZF, $\mu^2_G=1$ and there is no ICI, such that $\var(G) =0$, yielding  
$\SINR^{\rm{iid}\mathbb{C}}_{\rm ZF} =\frac{1}{ \mu_{|E|^2}}=\frac{\Nsc}{\sum_{k=0}^{\Nsc-1} |\widetilde{H}_k|^{-2}} 
$

For MMSE, we have $G_k = \frac{|\widetilde{H}_k|^2}{|\widetilde{H}_k|^2+1}$,  $\mu_{G}=\frac{1}{\Nsc} \sum_{k=0}^{\Nsc-1}
\frac{|\widetilde{H}_k|^2}{|\widetilde{H}_k|^2+1}$,
 $\mu_{G^2}=\frac{1}{\Nsc} \sum_{k=0}^{\Nsc-1} \frac{|\widetilde{H}_k|^4}{(|\widetilde{H}_k|^2+1)^2}$, and 
 $\mu_{|E|^2} =\frac{1}{\Nsc} \sum_{k=0}^{\Nsc-1}
\frac{|\widetilde{H}_k|^2}{(|\widetilde{H}_k|^2+1)^2}$. Thus
\begin{eqnarray}
\mu_{G^2}+\mu_{|E|^2}&=&   
\frac{1}{\Nsc} \sum_{k=0}^{\Nsc-1} \frac{|\widetilde{H}_k|^4-|\widetilde{H}_k|^2}{(|\widetilde{H}_k|^2+1)^2}\nonumber\\
&=&\frac{1}{\Nsc} \sum_{k=0}^{\Nsc-1} \frac{|\widetilde{H}_k|^2}{(|\widetilde{H}_k|^2+1)^2} =\mu_{G}.
\end{eqnarray}
Finally 
\begin{equation}
\expect{ | \ICI[m]|^2}+\sigma_n^2 =\mu_{G}(1-\mu_{G}). 
\end{equation}

\subsection{Proof of Lem.~\ref{lem:SINR_BPSK} (SINR of $\pi/2$-BPSK)}
\label{app:SINR_BPSK}

As a complex Gaussian random variable, the receiver noise is assumed circularly-symmetric, such that $e^{-\jrm \frac{\pi}{4}}  n[m]$ or  $ e^{-\jrm \frac{\pi}{2}\left(\frac{1}{2}+ (m  \ {\rm mod}\ 2)\right)}  n[m] $ have the same statistics as $n[m]$. Therefore, the variance of $ n_{\rm R}[m]$, i.e.,  the real part of the noise, is half that of $n[m]$. 

The BPSK symbols are assumed i.i.d., and the interference power is given by 
\begin{equation} \label{eqA:ICIbpsk}
\zeta^2_G  \triangleq \expect{ (\ICI_{\rm R}[m])^2} = \sum_{\substack{n=0\\n\neq m } }^{\Nsc-1}\Re \left\{ g_{m-n}\right\}^2.
\end{equation}

Using~\eqref{eq:gm} and trigonometric identities, 
\begin{multline}
\Re \left\{ g_{m-n}\right\}^2 =  \frac{1}{2\Nsc^2}   \sum_{k=0}^{\Nsc-1}  \sum_{h=0}^{\Nsc-1} G_k G_h \\
\left( \cos  \frac{2\pi(k+h)(m-n) }{\Nsc}    +\cos \frac{2\pi(k-h)(m-n)}{\Nsc} \right).
\label{eq:RgmnBPSK}
\end{multline} 

Now, inserting the above into~\eqref{eqA:ICIbpsk} and noting that
\begin{multline}
\sum_{\substack{n=0\\n\neq m } }^{\Nsc-1} \cos \frac{2\pi(k \pm h)(m-n)}{\Nsc} = 
\Re \left\{ \sum_{\substack{n=0\\n\neq m } }^{\Nsc-1} e^{\jrm \frac{2\pi(k \pm h)(m-n)}{\Nsc}} \right\} \\
\mkern-85mu = \begin{cases} 
\Nsc-1 & \text{ if } (k \pm h) = 0 \pmod \Nsc \\
-1 & \text{otherwise} 
\end{cases}.\mkern40mu
\end{multline}
\eqref{eqA:ICIbpsk} simplifies to~\eqref{eq:zeta} as
\begin{multline}
\zeta^2_G  = 
\frac{1}{2\Nsc^2}   \sum_{k=0}^{\Nsc-1}  G_k 
 \Bigg((\Nsc-1)G_{(-k)_{\mathrm{mod}\, \Nsc}}  \\  
\mkern120mu -\!\!\!\!\!\!\!\!\sum_{ \substack{h=0 \\ h\neq -k (\mathrm{mod}\,\Nsc)}}^{\Nsc-1} \!\!\!\!\!\!\!\! G_h  +  (\Nsc-1)G_{k} - \sum_{\substack{l=0\\ l\neq k}}^{\Nsc-1}G_l
\Bigg)\\
= \frac{1}{2\Nsc^2}  \!\!\!\! \sum_{k=0}^{\Nsc-1}  G_k 
\left( \Nsc\left(G_{(-k)_{\mathrm{mod}\, \Nsc}}    + G_{k}\right)- 2 \sum_{h=0}^{\Nsc-1}  G_h   \right)\\
=  \frac{1}{2\Nsc}  \!\!\!\! \sum_{k=0}^{\Nsc-1}  G_k 
\left(G_{(-k)_{\mathrm{mod}\, \Nsc}}    + G_{k}\right) - \left(\frac{1}{\Nsc}  \sum_{h=0}^{\Nsc-1}  G_h \right)^2 \!\!\!\!. 
\end{multline}

For the case of $\pi/2$-BPSK, 
\begin{equation} \label{eqA:ICIpi2bpsk}
\zeta^2_G  \triangleq \expect{ (\ICI_{\rm R}[m])^2} = \sum_{\substack{n=0\\n\neq m } }^{\Nsc-1}\Re \left\{ e^{\jrm \frac{\pi}{2}(n-m \ {\rm mod}\ 2 ) } g_{m-n}\right\}^2.
\end{equation}
and  
\begin{multline}
\Re \left\{ e^{\jrm(n-m \ {\rm mod}\ 2 ) } g_{m-n}\right\}^2 =  \frac{1}{2\Nsc^2}   \sum_{k=0}^{\Nsc-1}  \sum_{h=0}^{\Nsc-1} G_k G_h \\
\left( \cos  \frac{2\pi(k-h)(m-n) }{\Nsc}    +\cos\left(\frac{2\pi(k+h)}{\Nsc}-\pi\right)(m-n) \right).
\end{multline}
Compared with~\eqref{eq:RgmnBPSK}, the only difference lies in the $-\pi$ phase term in the second cosine.
Assuming $\Nsc$ even, and using the arithmetic sum formula, we have 
\begin{multline}
\sum_{\substack{n=0\\n\neq m } }^{\Nsc-1} \cos \left(\frac{2\pi(k+h)}{\Nsc}-\pi\right)(m-n) = 
\\
\mkern-85mu = \begin{cases} 
\Nsc-1 & \text{ if } (k + h) = \frac{\Nsc}{2} \pmod \Nsc \\
-1 & \text{otherwise} 
\end{cases}.
\end{multline}
and the final result follows similarly.

\section{RO-QPSK}
\subsection{Proof of Lem.~\ref{lem:PSD_RO-QPSK} (Spectrum)}
\label{App:PSD_RO-QPSK}

Directly from~\eqref{eq:X[k]}, we have
\begin{equation}
 \expect{X_k X^*_h} = \frac{1}{\Nsc} \sum_{m=0}^{\Nsc-1}
\sum_{l=0}^{\Nsc-1} \expect{x[m] x[l]^*} e^{-\jrm \frac{2\pi}{\Nsc} (km-hl)}.
\end{equation}

In the case of i.i.d. constellation symbols, $\expect{x[m]x[l]^*} = \delta_{m-l}$,  so that $ \expect{X_k X^*_h} = \frac{1}{\Nsc} \sum_{m=0}^{\Nsc-1}  e^{-\jrm \frac{2\pi}{\Nsc} m(k-h)}= \delta_{k-h} $. 

In the case of RO-QPSK, we have $\expect{|x[m]|^2} = 1$;  $\expect{x[m]x[l]^*} = -1/2$ for $l=m+1$ or   $l=m-1$; and zero otherwise. It follows that 
 \begin{eqnarray}
 \expect{X_k X^*_h} 
\!\!\!\!\!
&=&  \!\!\!\!\!   \frac{1}{\Nsc} \sum_{m=0}^{\Nsc-1} e^{-\jrm \frac{2\pi}{\Nsc} m (k-h)} \left( 1-\frac{ e^{\jrm \frac{2\pi}{\Nsc} h} +e^{-\jrm \frac{2\pi}{\Nsc} h} }{2} \right)
\nonumber\\
 \!\!\!\!\!&=& \!\!\!\!\!   \left( 1- \cos \frac{2\pi}{\Nsc} h  \right) \times \frac{1}{\Nsc} \sum_{m=0}^{\Nsc-1} e^{-\jrm \frac{2\pi}{\Nsc} m (k-h)} \nonumber\\
 \!\!\!\!\!&=& \!\!\!\!\!     \left( 1- \cos \frac{2\pi}{\Nsc} h  \right) \delta_{k-h}.
\end{eqnarray}

\subsection{Proof of Lem.~\ref{lem:PwNor_RO-QPSK} (Power Normalization)}
\label{App:PwNor_RO-QPSK}
By direct expansion of~\eqref{eq:s[n]}, we have 
\begin{equation}
\expect{|s[n]|^2} = \frac{\eta^2}{\Nfft} \sum_{k=0}^{\Nsc-1}
\sum_{h=0}^{\Nsc-1} F_k F_h^* \expect{X_k X^*_h} e^{\jrm \frac{2\pi}{\Nfft} n(k-h)}. 
\end{equation}

In the case of i.i.d. constellation symbols, $ \expect{X_k X^*_h}\delta_{k-h} $, and thus $\expect{|s[n]|^2} = \frac{\eta^2}{\Nfft} \sum_{k=0}^{\Nsc-1} |F_k|^2 = \frac{\Nsc}{\Nfft}$ with $\eta^2 = \Nsc / (\sum_{k=0}^{\Nsc-1} |F_k|^2) $.

In the case of RO-QPSK, using Lem.~\ref{lem:PSD_RO-QPSK} we have
\begin{equation}
\expect{|s[n]|^2} = \frac{\eta^2}{\Nfft} \sum_{k=0}^{\Nsc-1} |F_k|^2 w_k
\end{equation}
and thus the normalization must satisfy $\eta^2 = \Nsc / (\sum_{k=0}^{\Nsc-1} |F_k|^2 w_k) $. 

In the case of no FDSS, the normalization reduces to $\eta^2 = \Nsc / (\sum_{k=0}^{\Nsc-1} w_k) =1$ since
 \begin{eqnarray}
 \sum_{k=0}^{\Nsc-1} \left( 1- \cos \frac{2\pi}{\Nsc} k  \right)& =& \Re\left\{\sum_{k=0}^{\Nsc-1} \left( 1-e^{\frac{2\pi}{\Nsc} k} \right) \right\} \nonumber\\
& =&  \Nsc.
\end{eqnarray}

\subsection{Proof of Lem.~\ref{lem:RO-QPSK_Rxsig} (Received combined symbol)}
\label{app:nuEH} 

We start by separating the data and noise components in the received signal, rewriting~\eqref{eq:r[m]Mat} as   
$r[m] = d[m] + n[m]$
with 
\begin{equation} \label{eq:d[m]}
d[m] = \sum_{n=0}^{\Nsc-1}x[n] g_{m-n}. 
\end{equation}

Then the combined received signal~\eqref{eq:rtilde[l]} can rewritten as 
\begin{equation}
\tilde{r}[l]=  \tilde{d}[l] + \widetilde{\rm n}_l,
\end{equation}
where
$\tilde{d}[l]=  \tilde{d}_{\rm R} [l]+ 
\jrm \tilde{d}_{\rm I} [l]$, 
with 
\begin{eqnarray}
\tilde{d}_{\rm R} [l] &=&  \frac{1}{2}\Re \left\{   d[2l]-d[2l+1] \right\}
\\
\tilde{d}_{\rm I} [l] &=&\frac{1}{2} \Im \left\{  d[2l+1]-d[2l+2] \right\}.
\end{eqnarray}

By direct expansion of \eqref{eq:d[m]}  above, isolating the terms related to $q_l = \alpha_{2l} +\jrm \alpha_{2l+1}$, and noting that $g_{-a} = g_{a}^*$, and thus that $g_{a} +  g_{-a} = 2\Re\{ g_{a}\}$, it can be verified that  
\begin{eqnarray}
\tilde{d}_{\rm R} [l] &=& \mu_{w,G} \alpha_{2l}  + 
  \kappa \alpha_{2l+1}+ I_{{\rm R},l} \\
\tilde{d}_{\rm I} [l] &=& \mu_{w,G} \alpha_{2l+1} + 
 \kappa \alpha_{2l} +  I_{{\rm I},l} 
\end{eqnarray}
with
\begin{eqnarray}
\mu_{w,G} &=&  \Re\{g_0 -g_{-1}\}\\
\kappa   &=&  \frac{1}{2} \Im\{(2g_{1}-g_{2})\} 
\end{eqnarray} 
and $I_{{\rm R},l}$, $I_{{\rm I},l}$ represents the ICI terms independent of $\alpha_{2l}$ and $\alpha_{2l+1}$. 
Altogether, we can write
\begin{equation}
\tilde{d}[l] = \mu_{w,G} q_l +   \jrm \kappa  q_l^* + I_{{\rm R},l} + I_{{\rm I},l}
\end{equation}
and note that the component $ \jrm \kappa  q_l^*$  is also a form of ICI, as it corresponds to interference between I- and Q-branches. Hence, $\Ical_l = \jrm \kappa  q_l^* + I_{{\rm R},l} + I_{{\rm I},l} $.

Finally, by direct expansion of~\eqref{eq:gm} in $\mu_{w,G}$, we obtain~\eqref{eq:nuEH}.

\subsection{Proof of Prop.~\ref{lem:SINR_RO-QPSK} (SINR)}
\label{App:SINR_RO-QPSK}
From~\eqref{eq:rtilde[l]}, the SINR for RO-QPSK is defined as
\begin{equation}
\SINR^{\rm RO-QPSK} = \frac{\mu_{w,G}^2 }{\expect{|\Ical_l|^2}+ \expect{|\widetilde{\rm n}_l|^2}}.
\end{equation}


\subsubsection{Noise Power}
The combined noise components in the real and imaginary parts of~\eqref{eq:CombinedNoise}, both of the form $(n[m]-n[m+1])$,  are linear combinations of circularly symmetric Gaussian variables, and are therefore circularly symmetric Gaussian variables themselves, with equal power shared between real and imaginary parts.  Using $|1-e^{\jrm \frac{2\pi}{\Nsc}k} |^2= 2\left(1-\cos \frac{2\pi k}{\Nsc} \right) = 2 w_k$, the power of this combined noise  is  $
\expect{|n[m]-n[m+1]|^2} = \frac{2}{\Nsc} \sum_{k=0}^{\Nsc-1}w_k |E[k]|^2 $, 
which is independent of $m$. Hence, we have $\expect{|\widetilde{\rm n}_l|^2} =\frac{1}{4} \expect{|n[m]-n[m+1]|^2}  $, and the final result follows.

\subsubsection{Interference Power}
The interference power can be computed as
 \begin{eqnarray}
\expect{\left|\Ical_l\right|^2} &=& \expect{\left|\tilde{d}[l]-\mu_{w,G} q_l \right|^2} \nonumber\\
 &=& \expect{\left|\tilde{d}[l]\right|^2} + \mu_{w,G}^2 -
2\mu_{w,G}\expect{\Re\{\tilde{d}[l] q_l^*\}} \nonumber\\
 &=&\expect{\left|\tilde{d}[l]\right|^2}  -\mu_{w,G}^2 
+2\mu_{w,G}\kappa\Re\{ \jrm \expect{(q_l^*)^2}\}\nonumber\\
 &=& \nu_{w,G}  -\mu_{w,G}^2
\end{eqnarray}
 by  defining $
\nu_{w,G} \triangleq \expect{|\tilde{d}[l]|^2} $ (which will be verified to be independent of $l$) and observing that $\expect{(q_l^*)^2} =0$. 

Proceeding to compute the average power of $\tilde{d}[l]$, we have 
 \begin{equation} \label{eq:d2}
\nu_{w,G}  = 
 \expect{|\tilde{d}_{\rm R} [l]|^2}+ 
 \expect{|\tilde{d}_{\rm I} [l]|^2}
\end{equation}

We will derive in detail $\expect{|\tilde{d}_{\rm R} [l]|^2}$; by symmetry, similar derivations give $\expect{|\tilde{d}_{\rm I} [l]|^2}=\expect{|\tilde{d}_{\rm R} [l]|^2} $, and thus 
$\nu_{w,G} =  2 \expect{|\tilde{d}_{\rm R} [l]|^2}$.

Writing the real part as $\Re\{z\} = \frac{1}{2}(z+z^*)$, and denoting $m=2l$ for compactness, and  we expand
\begin{multline} \label{eq:dR2}
|\tilde{d}_{\rm R} [l]|^2 = \frac{1}{16}\big(
2|d[m]|^2+ 2|d[m+1]|^2 - 4\Re\{d[m]d[m+1]^*\} \\ 
 +2\Re\{d[m]^2\} + 2\Re\{d[m+1]^2\}  - 4\Re\{d[m]d[m+1]\} 
\big) .
\end{multline}

To derive the average of~\eqref{eq:dR2}, note that 
it involves two types of terms: 
the terms on the first line of the form $\expect{d[a]d[b]^*}$; and the terms in the second line  of the form $\expect{d[a]d[b]}$.

\paragraph{Terms of the forms $\expect{d[a]d[b]^*}$} 
For  integers $a,b$, we have
\begin{equation}
\expect{d[a]d[b]^*} =\sum_{n=0}^{\Nsc-1} \sum_{l=0}^{\Nsc-1} g_{a-n}g_{b-l}\expect{x[n]x[l]^*} .
\end{equation}
Here, the difference from QAM constellation case is that the transmitted constellation symbols are correlated. For each index $n$, $x[n]$ is  correlated only with $x[l]$ where $l=n-1$ or $l=n+1$, for which  $E[x[n]x[l]^* ]= -1/2$.
Hence, 
\begin{equation}
\expect{d[a]d[b]^*} =\sum_{n=0}^{\Nsc-1} 
 g_{a-n}( g_{b-n}^* - \frac{1}{2} g_{b-n-1}^*
-  \frac{1}{2} g_{b-n+1}^* ).
\end{equation}

From this, using \eqref{eq:PlancherelThG} and Euler formula, we get 
\begin{equation}
\expect{d[a]d[b]^*} = \frac{1}{\Nsc} 
\sum_{n=0}^{\Nsc-1} G_k^2  
e^{\frac{2\pi}{\Nsc}(a-b)k}
\left(1- \cos \frac{2\pi}{\Nsc}k\right).
\end{equation}

Applying the above to $(a,b)= (m,m),\, (m+1,m+1)$ and $(m,m+1)$, 
and combining them together, we  obtain 
\begin{multline} \label{eq:dR2_Line1}  
2|d[m]|^2+ 2|d[m+1]|^2 - 4\Re\{d[m]d[m+1]^*\}  \\
 =\frac{4}{\Nsc} 
\sum_{k=0}^{\Nsc-1} G_k^2  
\left(1- \cos \frac{2\pi}{\Nsc}k\right)^2 .
\end{multline}

\paragraph{Terms of the form $\expect{d[a]d[b]}$} 
For integers $a,b$, we have
\begin{equation}
\expect{d[a]d[b]} =\sum_{n=0}^{\Nsc-1} \sum_{l=0}^{\Nsc-1} g_{a-n}g_{b-l}\expect{x[n]x[l]} .
\end{equation}
The only non-zero correlated terms are $\expect{x[n]x[n+1]} = \frac{(-1)^{n+1}}{2}$ and  $\expect{x[n]x[n-1]} = \frac{(-1)^{n}}{2}$;  all other terms, including $\expect{x[n]^2} = 0$, are zero. 
Thus, 
\begin{equation}
\expect{d[a]d[b]} =\sum_{n=0}^{\Nsc-1} (-1)^n  g_{a-n}\frac{(g_{b-n+1}-g_{b-n-1})}{2}. 
\end{equation}

By expansion of~\eqref{eq:gm}, we have 
\begin{equation}
\frac{(g_{b-n+1}-g_{b-n-1})}{2} = \frac{\jrm}{\Nsc}\sum_{n=0}^{\Nsc-1} G_k e^{\jrm \frac{2 \pi}{\Nsc} k(b-n)} \sin \frac{2\pi}{\Nsc} k
\end{equation} 
and obtain
\begin{multline}
\expect{d[a]d[b]} = \frac{\jrm}{\Nsc^2} 
\sum_{k=0}^{\Nsc-1} \sum_{h=0}^{\Nsc-1} G_k G_h e^{\jrm \frac{2 \pi}{\Nsc} (ka+hb)} \\
\times \sin \left(\frac{2\pi}{\Nsc} h \right)
\sum_{n=0}^{\Nsc-1} (-1)^n e^{-\jrm \frac{2\pi}{\Nsc}n(k+h)}.
\end{multline}

Using the exponential sum formula, the term
$\sum_{n=0}^{\Nsc-1} (-1)^n e^{-\jrm \frac{2\pi}{\Nsc}n(k+h)}$ is zero in almost all cases except when $k+h=\frac{\Nsc}{2}\pmod \Nsc $, where it equals to $\Nsc$. After further simplification, 
\begin{multline}
\Re\{ \expect{d[a]d[b]}\} = 
\frac{(-1)^{b+1}}{\Nsc} 
\sum_{k=0}^{\Nsc-1} G_k G_{\left(\frac{\Nsc}{2}-k\right)}  \\
\times \sin \left( \frac{2 \pi}{\Nsc} k(a-b)\right)  \sin \left(\frac{2\pi}{\Nsc} k \right).
\end{multline}

Now recall that $m$ is even here, so we have $\Re\{ \expect{d[m]^2}\}=- \Re\{ \expect{d[m+1]^2}\}$ and these terms cancel out in~\eqref{eq:dR2}, while the last term becomes
\begin{multline} \label{eq:dR2_Line2}
\Re\{ \expect{d[m]d[m+1]}\} = 
\frac{-1}{\Nsc} 
\sum_{k=0}^{\Nsc-1} G_k G_{\left(\frac{\Nsc}{2}-k\right)}  \\
\times   \sin^2 \left(\frac{2\pi}{\Nsc} k \right).
\end{multline}

Finally, substituting~\eqref{eq:dR2_Line1} and ~\eqref{eq:dR2_Line1}
into~\eqref{eq:dR2} and then into \eqref{eq:d2}, we obtain

\begin{multline}
\nu_{w,G} = \frac{1}{2\Nsc} 
\sum_{k=0}^{\Nsc-1}G_k \Bigg( G_k
\left(1- \cos \frac{2\pi}{\Nsc}k\right)^2 \\
+G_{(\frac{\Nsc}{2}-k)}   \sin^2 \left(\frac{2\pi}{\Nsc} k \right) \Bigg).
\end{multline}

Observe that for $k \approx \frac{\Nsc}{4}$ and $k \approx \frac{3\Nsc}{4}$, the channel coefficients $G_k$ and $G_{\scriptscriptstyle \left(\frac{\Nsc}{2}-k\right)}$ are in close proximity, such that likely 
$G_k \approx G_{\scriptscriptstyle \left(\frac{\Nsc}{2}-k\right)}$.  Moreover, for coefficients $\frac{\Nsc}{4} \leq k \leq \frac{3\Nsc}{4}$, we have $\sin^2 \left(\frac{2\pi}{\Nsc} k \right) \leq \left(1- \cos \frac{2\pi}{\Nsc}k\right)^2 $, with $\sin^2 \left(\frac{2\pi}{\Nsc} k \right) = 0$ for $k=\frac{\Nsc}{2}$. Therefore, assuming this approximation for these coefficients has little impact. From this, we obtain 
\begin{equation}
\nu_{w,G} \approx
\frac{1}{\Nsc} 
\sum_{k=0}^{\Nsc-1}
\left(1- \cos \frac{2\pi}{\Nsc}k\right) G_k^2 .
\end{equation}

\section{Mutual Information}
\label{app:MutInf}
Consider a binary-input AWGN channel, $y  = h x + n$, where $x\in \Xcal = \{\pm a \}$  are antipodal and equiprobable symbols, $h$ is a real channel coefficient, and $n$ is a zero-mean real Gaussian noise with variance $\sigma_n^2$. 
With the assumption that $h$ is known at the receiver, the considered channel is $x \to (y,h)$, with mutual information $I(x;y,h) = \mathbb{E}_{h}\left[I(x;y|h)\right]$~\cite{telatar1999capacity}. 
Given $h$, the likelihood that symbol $x$ was transmitted is $p(y|x) = \frac{1}{\sigma_n \sqrt{2\pi}}e^{-\frac{(y-h x)^2}{2 \sigma_n^2}}$.  It follows that the mutual information of $(x,y)$ is
\begin{eqnarray}
\!\!\!\!\!\!I(x;y|h) &=& \mathbb{E}_{x,y}\left[ \log_2 \frac{p(x,y)}{p(x)p(y)} \right] \nonumber \\
 &=& \mathbb{E}_{x,y}\left[ \log_2 \frac{p(y|x)}{\sum_{\bar{x} \in \Xcal} p(y|\bar{x})p(\bar{x})} \right] \nonumber\\
 &=& 1- E_{x,y} \left[ \log_2 \frac{\sum_{\bar{x}\in \Xcal }p(y|\bar{x})}{p(y|x)}  \right] \nonumber\\
 &=& 1- E_{x,y} \left[\log_2 \frac{e^{-\frac{(y+ha)^2}{2 \sigma_n^2}}+e^{-\frac{(y-ha)^2}{2 \sigma_n^2}}}{e^{-\frac{(y-hx)^2}{2 \sigma_n^2}}}  \right]\!\!\!. \label{eq:I(x;y)}  
\end{eqnarray} 
 %
From this, $I(x;y,h)$ can be obtained by Monte Carlo evaluation over multiple transmission and channel realizations. Alternatively, \eqref{eq:I(x;y)} can be further simplified and evaluated via numerical integration, see for example~\cite[Ch. 3]{durisi20-11a}.

Treating the interference as Gaussian, the mutual information~\eqref{eq:I(x;y)} can be directly applied to~\eqref{eq:rR[m]} for $\pi/2$-BPSK with $\Xcal = \{\pm 1\}$,  $h=\mu_{G}$, and noise power  $\sigma_n^2 = \frac{\mu_{G}^2}{\SINR^{\rm BPSK}}$. 
By extension, for Gray-mapped QPSK, the I- and Q- branches are independently decoded, and the  mutual information is the sum of the two independent branches. That is, taking $y$ as the real or imaginary part of~\eqref{eq:r[m]}, ~\eqref{eq:I(x;y)} is applied with  $\Xcal =\{\pm \frac{1}{\sqrt{2}} \}$, $h=\mu_{G}$ and  $\sigma_n^2 = \frac{\mu_{G}^2}{2\SINR^{\rm{iid}\mathbb{C}}}$.

For RO-QPSK, the mutual information is computed in the same way as for QPSK but using the transmission equation~\eqref{eq:rtilde2[l]}, where $h= \mu_{w,G}$, and the real noise power on each I- and Q- branch is  $\sigma_n^2 = \frac{\mu_{G}^2}{2\SINR^{\rm RO-QPSK}}$. Finally, the mutual information is halved to account for the repetition coding, thereby aligning the effective number of channel uses.

\end{appendices} 
\bibliographystyle{IEEEtran}
\bibliography{RO-QPSK}

@Misc{3GPP_TR_38811,
  author       = {{3GPP TR 38.811}},
  howpublished = {V15.4.0},
  month        = sep,
  title        = {Study on {New Radio (NR)} to support non-terrestrial networks},
  year         = {2020},
}

@Misc{3GPP_TR_38901,
  author       = {{3GPP TR 38.901}},
  howpublished = {V16.1.0},
  month        = dec,
  title        = {Study on channel model for frequencies from 0.5 to 100 {GHz}},
  year         = {2019},
}

@Misc{3GPP_TS_38214,
  author       = {{3GPP TR 38.214}},
  howpublished = {V17.6.0},
  month        = {Jul.},
  title        = {{NR}; Physical layer procedures for data, Release 17},
  year         = {2023},
}

@Misc{3GPP_TR_38821,
  author       = {{3GPP TR 38.821}},
  howpublished = {V16.1.0},
  month        = may,
  title        = {Solutions for {NR} to support non-terrestrial networks ({NTN})},
  year         = {2021},
}

@Misc{3gpp.ts.38.101-5.17.1.0,
    author = {{3GPP TS 38.101-5}},
		  howpublished = {V17.1.0},
    title = {{NR }; User Equipment  {(UE)} radio transmission and reception; Part 5: Satellite access Radio Frequency ({RF}) and performance requirements},
    year = {2022},
    month = {Oct.}
}

@misc{3gpp.38.211.16.6.0,
    author = {{3GPP TS 38.211}},
    title = {{5G; NR;  Physical channels and modulation}},
	howpublished = {V16.6.0},
    year = {2021},
    month = {August}
}

@Article{Panaitopol22,
  author  = {D. Panaitopol and Y. Jin and R. Tang and C. Park},
  journal = {Int. J. Satell. Commun. Netw.},
  title   = {Requirements on Satellite Access Node and User Equipment for Non-Terrestrial Networks in {5G New Radio} of {3GPP} Release-17},
  year    = {2023},
  month   = {Aug.},
  number  = {3},
  pages   = {289-301},
  volume  = {41},
}

@article{SaarnisaariL21,
  author       = {Harri Saarnisaari and
                  Carlos H. M. de Lima},
  title        = {Application of {5G} new radio for satellite links with low peak-to-average
                  power ratios},
  journal      = {Int. J. Satell. Commun. Netw.},
  volume       = {39},
  number       = {4},
  pages        = {445--454},
  year         = {2021},
}

@article{PositionSurvey22,
  author       = {Fabricio dos Santos Prol et al.},
  title        = {Position, Navigation, and Timing {(PNT)} Through Low Earth Orbit {(LEO)}
                  Satellites: {A} Survey on Current Status, Challenges, and Opportunities},
  journal      = {{IEEE} Access},
  volume       = {10},
  pages        = {83971--84002},
  year         = {2022}
}

@ARTICLE{Nokia21,
  author={Nasarre, Ismael Peruga and Levanen, Toni and Pajukoski, Kari and Lehti, Arto and Tiirola, Esa and Valkama, Mikko},
  journal={IEEE Open J. Commun. Soc.}, 
  title={Enhanced Uplink Coverage for {5G NR}: Frequency-Domain Spectral Shaping With Spectral Extension}, 
  year={2021},
  volume={2},
  number={},
  pages={1188-1204}}

@INPROCEEDINGS{Nisar2007,
  author={Nisar, Muhammad Danish and Nottensteiner, Hans and Hindelang, Thomas},
  booktitle={Proc. IST Mobile Wireless Commun. Summit}, 
  title={On Performance Limits of {DFT} Spread {OFDM} Systems}, 
  year={2007},
  volume={},
  number={},
  pages={1-4}}

@ARTICLE{KimTVT18,
  author={Kim, Jubum and Yun, Yeo Hun and Kim, Chanhong and Cho, Joon Ho},
  journal={IEEE Trans.  Veh. Tech.}, 
  title={Minimization of {PAPR} for {DFT}-Spread {OFDM} With {BPSK} Symbols}, 
  year={2018},
  volume={67},
  number={12},
  pages={11746-11758}}

@INPROCEEDINGS{TENCON18,
  author={Kim, Jubum and Yun, Yeo Hun and Kim, Chanhong and Cho, Joon Ho},
  booktitle={Proc. IEEE Region 10 Int. Conf. TENCON}, 
  title={{PAPR} Reduction by Constellation Rotation and Pulse Shaping for {DFT}-Spread {OFDM} with {QPSK} Symbols}, 
  year={2018},
  volume={},
  number={},
  pages={0090-0095}}

@INPROCEEDINGS{ErkaiPIMRC24,
  author={Chen, Erkai and Pitaval, Renaud-Alexandre and Popovi\'c, Branislav M. and Qin, Yi},
  booktitle={Proc.  IEEE Int. Symp. Pers. Indoor Mob. Radio Commun. }, 
  title={Direct Satellite Access Using Multi-Dimensional Constellations}, 
  year={2024},
  volume={},
  number={},
  pages={1-7}}

@ARTICLE{VanDeBeekComLet09,
  author={Van De Beek, Jaap},
  journal={IEEE Commun. Let.}, 
  title={Sculpting the multicarrier spectrum: a novel projection precoder}, 
  year={2009},
  volume={13},
  number={12},
  pages={881-883}}

@ARTICLE{TCOM25Pitaval,
  author={Pitaval, Renaud-Alexandre and Tie, Xiaolei},
  journal={IEEE Trans. Commun.}, 
  title={{DFT-s-OFDM}-based On-Off Keying for Low-Power Wake-Up Signal}, 
  year={2025},
  volume={73},
  number={10},
  pages={1-17}}

@ARTICLE{PitavalFDSS-SE,
  author={Renaud-Alexandre Pitaval and Fredrik Berggren and Branislav M. Popovic},
  journal={IEEE Trans. Veh. Technol. (early access)}, 
  title={Optimum Spectrum Extension for {PAPR} Reduction of {DFT-s-OFDM}}, 
  year={2026}, 
	 pages={1-17}}

@ARTICLE{BerggrenComLet25,
  author={Berggren, Fredrik and Popovi\'c, Branislav M.},
  journal={IEEE Commun. Let.}, 
  title={{GNSS}-Independent Uplink {LEO} Satellite Synchronization}, 
  year={2025},
  volume={29},
  number={6},
  pages={1245-1249}}

@misc{R1-2506612,
author={{\relax 3GPP TSG RAN WG1 -- Moderator (Thales)} },
title={	{FL} Summary \#4 - {NR-NTN GNSS} resilience},
howpublished = {Meeting  122, document R1-2506612},
year = 2025,
month ={Aug.}
}

@misc{R1-2506550,
author={{\relax 3GPP TSG RAN WG1 -- Moderator (Nokia)} },
title={	 Feature Lead summary \#1 on {6GR} waveform},
howpublished = {Meeting  122, document R1-2506550},
year = 2025,
month ={Aug.}
}

@ARTICLE{SaeedSurvey20,
  author={Saeed, Nasir and Elzanaty, Ahmed and Almorad, Heba and Dahrouj, Hayssam and Al-Naffouri, Tareq Y. and Alouini, Mohamed-Slim},
  journal={IEEE Commun. Surveys  Tutorials}, 
  title={CubeSat Communications: Recent Advances and Future Challenges}, 
  year={2020},
  volume={22},
  number={3},
  pages={1839-1862}}

@ARTICLE{FeherTVT07,
  author={Park, Hyung Chul and Lee, Kwyro and Feher, Kamilo},
  journal={IEEE Trans. Veh. Tech.}, 
  title={Continuous Phase Modulation of {F-QPSK-B} Signals}, 
  year={2007},
  volume={56},
  number={1},
  pages={157-172}}

@ARTICLE{RahmatallahSurvey13,
  author={Rahmatallah, Yasir and Mohan, Seshadri},
  journal={IEEE Commun. Surveys Tutorials}, 
  title={Peak-To-Average Power Ratio Reduction in {OFDM} Systems: A Survey And Taxonomy}, 
  year={2013},
  volume={15},
  number={4},
  pages={1567-1592}}

@ARTICLE{ThompsonTCOM08,
  author={Thompson, Steve C. and Ahmed, Ahsen U. and Proakis, John G. and Zeidler, James R. and Geile, Michael J.},
  journal={IEEE Trans. Commun.}, 
  title={Constant Envelope {OFDM}}, 
  year={2008},
  volume={56},
  number={8},
  pages={1300-1312}}

@incollection{Bölcskei2003,
  author    = {Helmut Bölcskei},
  title     = {{Orthogonal Frequency Division Multiplexing based on Offset QAM}},
  booktitle = {{Advances in Gabor Analysis}},
  editor    = {{Feichtinger, H. G. and Strohmer, T.}},
  publisher = {{Birkhäuser}},
  pages     = {{321--352}},
  year      = {{2003}}}

@ARTICLE{AulinTCOM81,
  author={Aulin, T. and Sundberg, C.},
  journal={IEEE Trans. Commun.}, 
  title={Continuous Phase Modulation - Part {I}: Full Response Signaling}, 
  year={1981},
  volume={29},
  number={3},
  pages={196-209}}

@ARTICLE{MurotaTCOM81,
  author={Murota, K. and Hirade, K.},
  journal={IEEE Trans. Commun.}, 
  title={{GMSK} Modulation for Digital Mobile Radio Telephony}, 
  year={1981},
  volume={29},
  number={7},
  pages={1044-1050}}

@ARTICLE{MajeedTCOM97,
  author={Majeed, R.M. and McLane, P.J.},
  journal={IEEE Trans. Commun}, 
  title={Modulation techniques for on-board processing satellite communications}, 
  year={1997},
  volume={45},
  number={12},
  pages={1508-1512}}

@ARTICLE{BoydTCOM19,
  author={Boyd, Christopher and Pitaval, Renaud-Alexandre and Tirkkonen, Olav and Wichman, Risto},
  journal={IEEE Trans. Commun}, 
  title={Time–Frequency Localization Measures for Packets of Orthogonally Multiplexed Signals}, 
  year={2019},
  volume={67},
  number={9},
  pages={6374-6385}}

@book{durisi20-11a,
  author = {Giuseppe Durisi and Alejandro Lancho},
  month = {Nov. },
  title = {Transmitting short packets over wireless channels---an information-theoretic perspective},
  url = {https://gdurisi.github.io/fbl-notes/},
  year = {2020}
}

@article{telatar1999capacity, 
title={Capacity of multi-antenna {G}aussian channels}, 
author={Telatar, I Emre}, journal={ Eur. Trans. Telecommun.. }, volume={10}, number={6}, pages={585--596}, year={1999}, publisher={Wiley-Blackwell} }

@misc{R1-2506218,
author={{\relax 3GPP TSG RAN WG1 -- Qualcomm Incorporated} },
title={Waveforms for {6GR}},
howpublished = {Meeting  122, document R1-2506218},
year = 2025,
month ={Aug.}
}

@misc{R1-2505913,
author={{\relax 3GPP TSG RAN WG1 -- Apple} },
title={Waveforms for {6GR} air interface},
howpublished = {Meeting  122, document R1-2505913},
year = 2025,
month ={Aug.}
}

@misc{R1-1705060,
author={{\relax 3GPP TSG RAN WG1 -- Huawei, HiSilicon} },
title={Performance evaluation for pi/2 BPSK with FDSS },
howpublished = {Meeting  88bis, document R1-1705060},
year = 2017,
month ={Apr.}
}

@misc{R1-260XXXX,
author={{\relax 3GPP TSG RAN WG1} },
title={Chair notes},
howpublished = {Meeting  124bis},
year = 2026,
month ={Feb.}
}

@misc{orellana2026capacityboundsdopplerofdm,
      title={Capacity Bounds on {Doppler OFDM} Channels}, 
      author={Pablo Orellana and Zheng Li and Jean-Marc Kelif and Sheng Yang and Shlomo Shamai},
      year={2026},
      eprint={2602.04862},
      archivePrefix={arXiv},
      primaryClass={cs.IT},
      url={https://arxiv.org/abs/2602.04862}}

@ARTICLE{SanchezTVT2011,
  author={Sanchez-Sanchez, Juan J. and Aguayo-Torres, Mari Carmen and Fernandez-Plazaola, Unai},
  journal={IEEE Trans.  Veh. Technol.}, 
  title={BER Analysis for Zero-Forcing {SC-FDMA} Over {N}akagami-m Fading Channels}, 
  year={2011},
  volume={60},
  number={8},
  pages={4077-4081}}

@INPROCEEDINGS{SanchezVTC09,
  author={Sanchez-Sanchez, Juan J. and Fernandez Plazaola, Unai and Aguayo-Torres, M. C. and Entrambasaguas, J. T.},
  booktitle={Proc. IEEE Veh. Technol. Conf. Fall}, 
  title={Closed-Form BER Expression for Interleaved {SC-FDMA} with {M-QAM}}, 
  year={2009},
  pages={1-5}}

@ARTICLE{SanchezTVT2013,
  author={Sanchez-Sanchez, Juan J. and Aguayo-Torres, Mari Carmen and Fernandez-Plazaola, Unai},
  journal={IEEE Trans.  Veh. Technol.}, 
  title={Spectral Efficiency of Interleaved {SC-FDMA} With Adaptive Modulation and Coding Over {N}akagami-m Fading Channels}, 
  year={2013},
  volume={62},
  number={8},
  pages={3663-3670}}

@ARTICLE{SahinComLet2021,
  author={Şahin, Alphan and Hosseini, Nozhan and Jamal, Hosseinali and Hoque, Safi Shams Muhtasimul and Matolak, David W.},
  journal={IEEE Commun. Lett.}, 
  title={{DFT}-Spread-{OFDM}-Based Chirp Transmission}, 
  year={2021},
  volume={25},
  number={3},
  pages={902-906}}

@ARTICLE{WLMMSETWC24,
  author={Chae, Joohee and Choi, Jeonghoon and Kim, Jubum and Cho, Joon Ho},
  journal={IEEE Trans.  Wireless Commun.}, 
  title={A Low-Complexity Widely-Linear {MMSE} Equalizer for {DFT}-Spread {OFDM} With Frequency-Domain Spectrum Shaping}, 
  year={2024},
  volume={23},
  number={4},
  pages={3465-3479}}

@ARTICLE{WLNyquistTCOM21,
  author={Choi, Jeonghoon and Kim, Jubum and Cho, Joon Ho and Lehnert, James S.},
  journal={IEEE Trans. Commun.}, 
  title={Widely-Linear {N}yquist Criteria for {DFT}-Spread {OFDM} of Constellation-Rotated {PAM} Symbols}, 
  year={2021},
  volume={69},
  number={5},
  pages={2909-2922}}

@INPROCEEDINGS{QC_WiSEE23,
  author={Hassan, Mohamad Sayed and Saha, Chiranjib and Lianghai, Ji and Alvarino, Alberto Rico and Ma, Jun and Liu, Le and Wu, Qiang},
  booktitle={Proc. IEEE International Conference on Wireless for Space and Extreme Environments (WiSEE)}, 
  title={{NTN: from 5G NR to 6G}}, 
  year={2023},
  volume={},
  number={},
  pages={173-178}}

@inproceedings{Rapp1991HPA,
  title={Effects of {HPA}-Nonlinearity on a {4-DPSK/OFDM-Signal} for a Digital Sound Broadcasting System},
  author={Rapp, Christoph},
  booktitle={Proc. Eur. Conf. Satell. Commun.},
  pages={179--184},
  year={1991},
  address={Liege, Belgium},
  month={Oct.}
}

@article{hossain2020dft,
  title={{DFT}-Spread {OTFS} communication system with the reductions of {PAPR} and nonlinear degradation},
  author={Hossain, M. N. and Sugiura, Y. and Shimamura, T. and Ryu, H. G.},
  journal={Wirel. Pers. Commun.},
  volume={115},
  number={3},
  pages={2211--2228},
  year={2020},
  publisher={Springer}
}

@article{hossain2019nonlinear,
  title={Nonlinear Characteristics of {DFT}-Spread {WR-OFDM} System for Spectrum-efficient Communications},
  author={Hossain, Md. Najmul and Shimamura, Tetsuya and Ryu, Heung-Gyoon},
  journal={IEIE Trans. Smart Process. Comput.},
  volume={8},
  number={6},
  pages={499--507},
  year={2019},
  publisher={The Institute of Electronics and Information Engineers}
}


\end{document}